\documentclass{aa}
\usepackage{graphicx}
\usepackage{subfigure}
\usepackage{txfonts}
\usepackage{natbib}  
\usepackage{color}
\usepackage{ulem}
\usepackage{amsmath}  
\begin{document}

   \title{Sodium abundances of AGB and RGB stars in Galactic globular clusters}

   \subtitle{II. Analysis and results of NGC\,104, NGC\,6121, and NGC\,6809 
            \thanks{Based on observations made with ESO telescopes at the La Silla Paranal Observatory under programme ID 093.D-0818(A).}
            \thanks{Full Tables 3, 5, 7 are only available in electronic form at the CDS via anonymous ftp to cdsarc.u-strasbg.fr (130.79.128.5) 
                    or via http://cdsweb.u-strasbg.fr/cgi-bin/qcat?J/A+A/.}}

   \author{Y. Wang 
           \inst{1,2,3}
           \and
           F. Primas \inst{3}
           \and
           C.Charbonnel \inst{4,5} 
           \and
           M. Van der Swaelmen \inst{6} 
           \and
           G. Bono \inst{7,8}
           \and
           W. Chantereau \inst{4}
           \and
           G. Zhao \inst{1,2}
          }

   \institute{Key Laboratory of Optical Astronomy, National Astronomical Observatories, Chinese Academy of Sciences, Beijing 100012, PR China \\
              \email{ywang@nao.cas.cn}
         \and
             School of Astronomy and Space Science, University of Chinese Academy of Sciences, Beijing 100049, China \\
         \and
             European Southern Observatory (ESO), Karl-Schwarschild-Str. 2, 85748, Garching b. München, Germany \\
              \email{fprimas@eso.org}
         \and
             Department of Astronomy, University of Geneva, Chemin des Maillettes 51, 1290, Versoix, Switzerland \\
              \email{corinne.charbonnel@unige.ch}
         \and
             IRAP, UMR 5277 CNRS and Université de Toulouse, 14 Av. E. Belin, F-31400 Toulouse, France \\
         \and
             Institut d’Astronomie et d’Astrophysique, Université Libre de Bruxelles, CP. 226, Boulevard du Triomphe, 1050 Brussels, Belgium \\
         \and
             Dipartimento di Fisica, Università di Roma Tor Vergata, via della Ricerca Scientifica 1, 00133 Rome, Italy \\
         \and
             INAF – Osservatorio Astronomico di Roma, via Frascati 33, Monte Porzio Catone, Rome, Italy \\
             }
             

 
  \abstract
   {}
   {We investigate the Na abundance distribution of asymptotic giant branch (AGB) stars in Galactic globular clusters (GCs) and its possible 
    dependence on GC global properties, especially age and metallicity.} 
   {We analyze high-resolution spectra of a large sample of AGB and red giant branch (RGB) stars in the Galactic GCs NGC\,104, NGC\,6121, and 
    NGC\,6809 obtained with FLAMES/GIRAFFE at ESO/VLT, and determine their Na abundances. This is the first time that the AGB stars in NGC\,6809 
    are targeted. Moreover, to investigate the dependence of AGB Na abundance dispersion on GC parameters, we compare the AGB [Na/H] distributions 
    of a total of nine GCs, with five determined by ourselves with homogeneous method and four from literature, covering a wide range of GC parameters.}
   {NGC\,104 and NGC\,6809 have comparable AGB and RGB Na abundance distributions revealed by the K$-$S test, while NGC\,6121 shows a lack of 
    very Na-rich AGB stars. By analyzing all nine GCs, we find that the Na abundances and multiple populations of AGB stars form complex 
    picture. In some GCs, AGB stars have similar Na abundances and/or second-population fractions as their RGB counterparts, while some GCs 
    do not have Na-rich second-population AGB stars, and various cases exist between the two extremes. 
    In addition, the fitted relations between fractions of the AGB second population and GC global parameters show that the AGB second-population 
    fraction slightly anticorrelates with GC central concentration, while no robust dependency can be confirmed with other GC parameters.}
   {Current data roughly support the prediction of the fast-rotating massive star (FRMS) scenario. However, considering the weak observational 
    and theoretical trends where scatter and exceptions exist, the fraction of second-population AGB stars can be affected by more than one or 
    two factors, and may even be a result of stochasticity.} 


   \maketitle
%

\section{Introduction}
\label{section:introduction}

Galactic globular clusters (GCs) have been the subject of a variety of stellar evolution studies; first, because for a long time they were thought to 
consist of a single stellar population (i.e., coeval and sharing the same initial chemical properties) thus making them the ideal stellar laboratory, 
and second, because of their  more recently discovered intriguing complexity of being inhabited by multiple stellar populations, a feature that has 
turned out to be common to most Galactic globular clusters. 

This multiplicity has been identified based on the appearance of multimodal sequences in different regions (e.g., main sequence (MS), sub-giant branch 
(SGB), red giant branch (RGB) and horizontal branch (HB)) of GC color-magnitude diagrams (CMD; e.g., \citealp{Piotto2012,Milone2012b,Piotto2015,Milone2015a,Milone2015b,Nardiello2015b}) 
that were associated to the variations in He and light element (e.g., C, N, and O) abundances in their initial chemical composition (see e.g., \citealp{Milone2012b,Chantereau2015} and references therein). 

With the advent of multi-object spectrographs mounted on 8$-$10m-class telescopes, detailed chemical abundance analyses have also uncovered specific 
features $-$ elemental (anti-)correlations $-$ between the light element pairs C$-$N, O$-$Na, Mg$-$Al (e.g., \citealp{Carretta2016} for a recent review).
These are commonly interpreted as a signature of the existence of at least two stellar populations: 
a first-population (1P) of GC stars displaying Na and O abundances consistent with that of halo field stars of similar metallicity; and a second-population 
(2P) of GC stars characterized by Na overabundances and O deficiencies. Although the O$-$Na pair is probably the most documented one in terms of data, a 
similar picture is also derived from the other pairs, Mg$-$Al (e.g., \citealp{Carretta2014a,Carretta2014b}) and C$-$N (e.g., \citealp{Carretta2005,Pancino2010}). 

A wealth of observational data has been collected and analyzed for a respectable number of Galactic GCs at different evolutionary phases, for example, 
from MS and SGB to RGB and HB (see \citealp{Wang2016}, here after Paper\,I, for a more detailed summary). 
However, asymptotic giant branch (AGB) stars have rarely been targeted in a systematic way until very recently, due to their paucity in GCs (a result 
of their short lifetime) and inefficient RGB/AGB separation criteria. Recently, several studies have focused on GC AGB stars, mainly stimulated by the 
claim by \citet[hereafter C13]{Campbell2013} that no Na-rich 2P AGB stars exists in NGC\,6752. 
This striking finding was challenged by  \citet{Lapenna2016} who re-observed the 20 AGB stars of the C13 sample at higher resolution with ESO-VLT/UVES, 
and found that both 1P and 2P stars populate the AGB of NGC 6752, with only stars with extreme Na enhancement missing. The presence of 2P AGB stars 
in this GC is also supported by \citet{Gruyters2017} who claimed a photometric split on the AGB sequence using Str{\"o}mgrem photometry.
Other GC AGB stars have also been scrutinized. \citet{Johnson2015} studied 35 AGB stars in 47\,Tuc (NGC\,104) and found that the AGB and RGB samples 
of 47\,Tuc have nearly identical [Na/Fe] dispersions, with only a small fraction ($\lesssim 20\%$) of Na-rich stars that may fail to ascend the AGB. 
\citet{GarciaHernandez2015} showed that 2P AGB stars exist in metal-poor GCs with a study of Mg and Al abundances in 44 AGB stars from four metal-poor 
GCs (M\,13, M\,5, M\,3 and M\,2). In Paper\,I, we looked at NGC\,2808 and also found that its AGB and RGB stars share similar Na abundance dispersions. 
Moreover, we found more Na-rich 2P stars in the AGB sample than in the RGB one. 
The multiple populations in AGB stars in NGC\,2808 was also confirmed by \citet{Marino2017} who carried out a study combining spectroscopy and photometry. 
They also looked at NGC\,6121 (M4) and found it hosts two main populations in agreement with the finding by \citet{Lardo2017} that AGB stars show broadened 
distribution in close analogy with their RGB counterparts in the $C_{UBI}$\footnote{$C_{UBI} = (U - B) - (B - I)$}$- V$ diagram. We note, however, that 
their conclusion on NGC\,6121 contradicts the result of \citet{MacLean2016} who found that the AGB is populated by Na-poor and O-rich stars (from the 
analysis of 15 AGB and 106 RGB stars). 
NGC\,6266 (M62) was also found to have only 1P AGB stars by \citet{Lapenna2015}, but their  conclusion may be affected by the small number statistics 
of their sample (6 AGB and 13 RGB stars).

It is now largely accepted that GCs experienced self-enrichment during their early evolution, and that 2P stars formed out of the Na-rich, O-poor 
ashes of high-temperature-burning hydrogen ejected by more massive 1P stars and diluted with interstellar gas (e.g., \citealp{PrantzosCharbonnel2006, 
Prantzos2007}). However, the nature of the polluters remains highly debated, as well as the mode and timeline of the formation of 2P stars. 
Among the most commonly-invoked scenarios, one finds fast-rotating massive stars (FRMS, with initial masses above $\sim25\,\mathrm{M_{\sun}}$; 
\citealp{MaederMeynet2006, PrantzosCharbonnel2006, Decressin2007b, Decressin2007a, Krause2013}), massive AGB stars (with initial masses of 
$\sim6-11\,\mathrm{M_{\sun}}$; \citealp{Ventura2001, DErcole2010, Ventura2011, Ventura2013}) and supermassive stars ($\sim10^{4}\,\mathrm{M_{\sun}}$; 
\citealp{Denissenkov2014,Denissenkov2015}). Other possible polluters have also been explored, like massive stars in close binaries ($10-20\,\mathrm{M_{\sun}}$; 
\citealp{deMink2009, Izzard2013}), FRMS paired with AGB stars \citep{Sills2010} or with high-mass interactive binaries \citep{Bastian2013, Cassisi2014a}.
So far, none of the proposed scenarios have been able to reconcile all aspects of the formation and evolution of GCs with the spectroscopic and photometric 
complexity exhibited by these systems, nor with the new constraints coming from extragalactic young massive clusters that have masses similar to the 
initial mass postulated for GCs within the self-enrichment framework (e.g., \citealp{Bastian2015, Renzini2015, Krause2016, Charbonnel2016EAS}). 

One key feature to pay attention to is how the various scenarios differ from one another. The origin and amount of He enrichment predicted for 
2P stars is one such example, which has important consequences on the way the various sequences of the CMDs can be populated (e.g., \citealp{DAntona2010, 
Chantereau2015,Chantereau2016}). Interestingly enough, the different theoretical predictions for the coupling between He and Na enrichments in the initial 
composition of 2P stars are expected to differentially affect the extent of the Na dispersion today among RGB and AGB stars in individual GCs, in proportions 
that depend on their age and metallicity \citep{CCWC2016}. 
In the original FRMS framework, 2P low-mass stars are predicted to be born with large and correlated spreads in both He and Na abundances \citep{Decressin2007b}. 
Since the lifetime and the fate of stars strongly depend on their initial He content, the FRMS scenario predicts that, above a certain threshold, or cutoff, 
of initial He and Na abundance, 2P stars do miss the AGB (so-called AGB-manqu\'{e}) and evolve directly towards the white dwarf stage after central He 
burning \citep{Charbonnel2013, Chantereau2015}. 
This provides in principle a nice explanation for the lack of Na-rich AGB stars observed in NGC\,6752 by C13. 
\citet{CCWC2016} have also shown that within the original FRMS scenario, the maximum Na content expected on the AGB is a (weak) function of both the 
metallicity and the age of GCs. Namely, at a given metallicity, younger clusters are expected to host AGB stars exhibiting a larger Na spread than older 
clusters, and at a given age, higher Na dispersion along the AGB is predicted in metal-poor GCs than in the metal-rich ones. Additionally, mass loss along 
the RGB has been shown to strongly impact the evolution of low-mass stars on the AGB, and  therefore to modify the theoretical Na cut on the AGB (\citealp{CCWC2016}; 
see also \citealp{Cassisi2014b}): 
the higher the mass loss, the stronger the trends with age and metallicity. However, the situation might be much more complex, as revealed by the derivation 
of the helium variations between 1P and 2P stars in several GCs by multiwavelength photometry of multiple sequences, which turn out to be much lower than 
predicted by both the original FRMS and AGB scenario \citep{Anderson2009, diCriscienzo2010, Pasquini2011, Milone2012, Milone2012b, Milone2013, Piotto2013, 
Marino2014, Larsen2015, Milone2015b, Nardiello2015a, Nardiello2015b}. 
It is therefore fundamental to test model predictions with observations of AGB and RGB stars in GCs spanning a large range in age and metal content. 
This is necessary to probe the degree of stochasticity lying behind the broad variety of chemical patterns observed in GCs (e.g., \citealp{Bastian2015}).

Considering the current limited sample of GC AGB stars with accurately determined Na (and O) abundances, we carried out a systematic observational 
campaign of four GCs (NGC\,104, NGC\,2808, NGC\,6121 and NGC\,6809). We have already presented our first results of NGC\,2808 in Paper\,I. Here, we 
report and discuss our results of the other three GCs, investigating whether the presence of Na-rich stars on AGB is dependent on metallicity and/or 
other GC parameters. 

The paper is organized as follows. In Sects. \ref{section:obsreduction} and \ref{section:stellarparametersabundances}, we describe the observations and 
detail the analysis of the data for our sample of GCs. In Sect. \ref{section:OtherGCs}, we present the re-analysis of C13 data for NGC\,6752 and we show 
other four GCs from the literature. In Sect. \ref{section:discussion}, we compare the behavior of Na along the AGB and the RGB for the full GC sample 
(ours plus literature); we discuss also the possible correlations between the corresponding fractions of 1P and 2P stars and the GC global properties, 
and compare with the theoretical predictions of the original FRMS scenario. A summary and concluding remarks close the paper in Sect. \ref{section:summary}. 


\begin{table*} \small
\caption{Summary of the observations.}    
\label{obslog}
\centering
\begin{tabular}{c c c c c c c c}
\hline\hline     
 Instrument   & Setup  &   R   & $\lambda$-range &  Exp.Time (s)       &  Exp.Time (s)    &  Exp.Time (s)       \\
              &         &       &   (nm)          &  NGC\,104 (47\,Tuc) & NGC\,6121 (M\,4) & NGC\,6809 (M\,55)   \\
\hline
 GIRAFFE      & HR\,13  & 22500 & 612.0$-$640.5   & 2$\times$1200                 & 1$\times$2700                 & 2$\times$3600                  \\
              & HR\,15  & 19300 & 660.7$-$696.5   & 2$\times$1200                 & 1$\times$1800                 & 2$\times$2770                  \\
              & HR\,19  & 14000 & 774.5$-$833.5   & 2$\times$2700                 & 2$\times$2700                 & 2$\times$2770                  \\
 UVES$-$fibre & Red 580 & 47000 & 480$-$680       & 2$\times$3600 + 4$\times$1200 & 3$\times$2700 + 1$\times$1800 & 2$\times$3600 + 4$\times$2770  \\
\hline
\end{tabular}
\end{table*}

\begin{table*} \small
\caption{Number of stars (observed and confirmed cluster members) and derived barycentric radial velocities.}   
\label{basic1}
\centering
\begin{tabular}{c c c c c c c}
\hline\hline   
  GC      & Observed stars &   AGB star   &   RGB star    &     RV      & $\sigma_\mathrm{RV}$ &     RV$_\mathrm{Harris}$         \\ 
           &             &    members    &    members     & ($\mathrm{km}\,\mathrm{s}^{-1}$) & ($\mathrm{km}\,\mathrm{s}^{-1}$) & ($\mathrm{km}\,\mathrm{s}^{-1}$)    \\ 
\hline 
 NGC\,104  &     94      &  46 + 4      &  40 + 4       &    -17.3    &   9.9   &    -18.0       \\
 NGC\,6121 &     95      &  17 + 2      &  63 + 5       &     70.4    &   3.3   &     70.7       \\
 NGC\,6809 &    110      &  23 + 1 (1)  &  74 + 10 (6)  &    173.6    &   3.7   &    174.7       \\
\noalign{\smallskip}
 NGC\,2808 &    100      &  30 + 3      &  38 + 2       &    104.6    &   8.0   &    101.6       \\
\hline    
\end{tabular}
\tablefoot{Numbers provided in 3$^{th}$ and 4$^{th}$ columns are given in the format of stars observed with ``GIRAFFE $+$ UVES'', 
           with the number of stars in common specified between parentheses. Our data on NGC\,2808 from Paper\,I are also reported to help in the comparison.
           }
\end{table*}


\section{Observation and data reduction}
\label{section:obsreduction}

\begin{figure}[ht]
   \centering
     \includegraphics[width=0.44\textwidth]{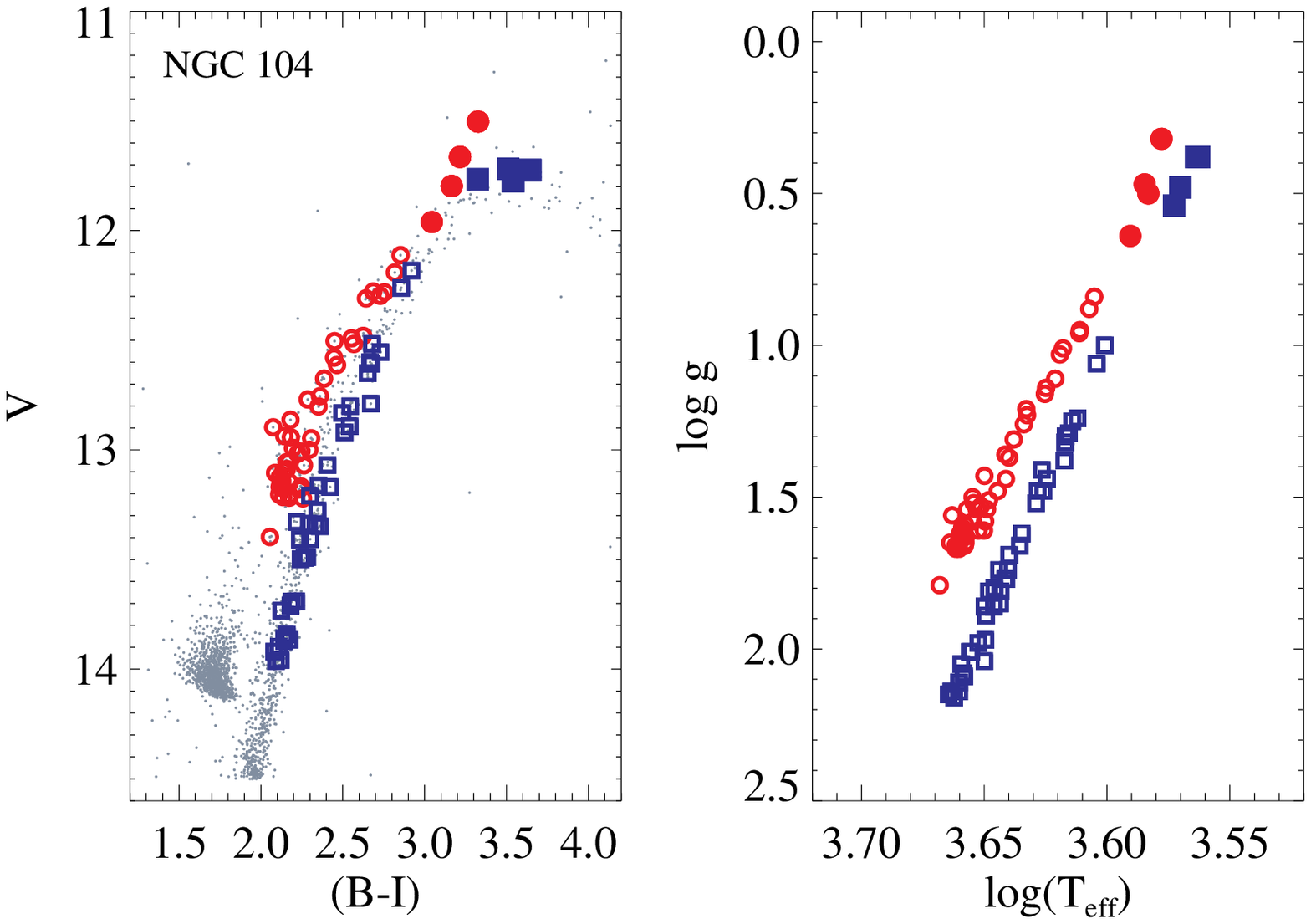}
     \includegraphics[width=0.44\textwidth]{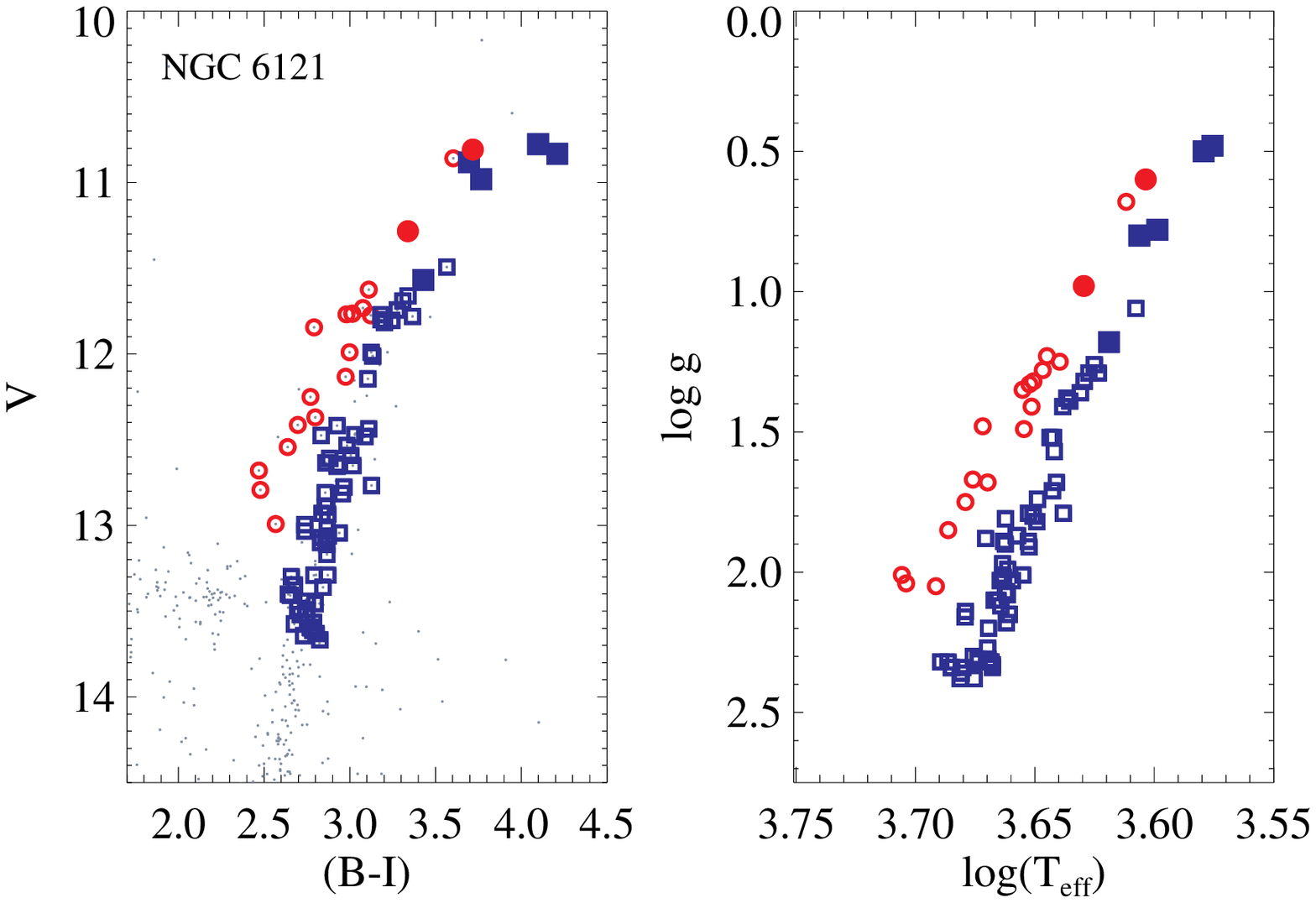}
     \includegraphics[width=0.44\textwidth]{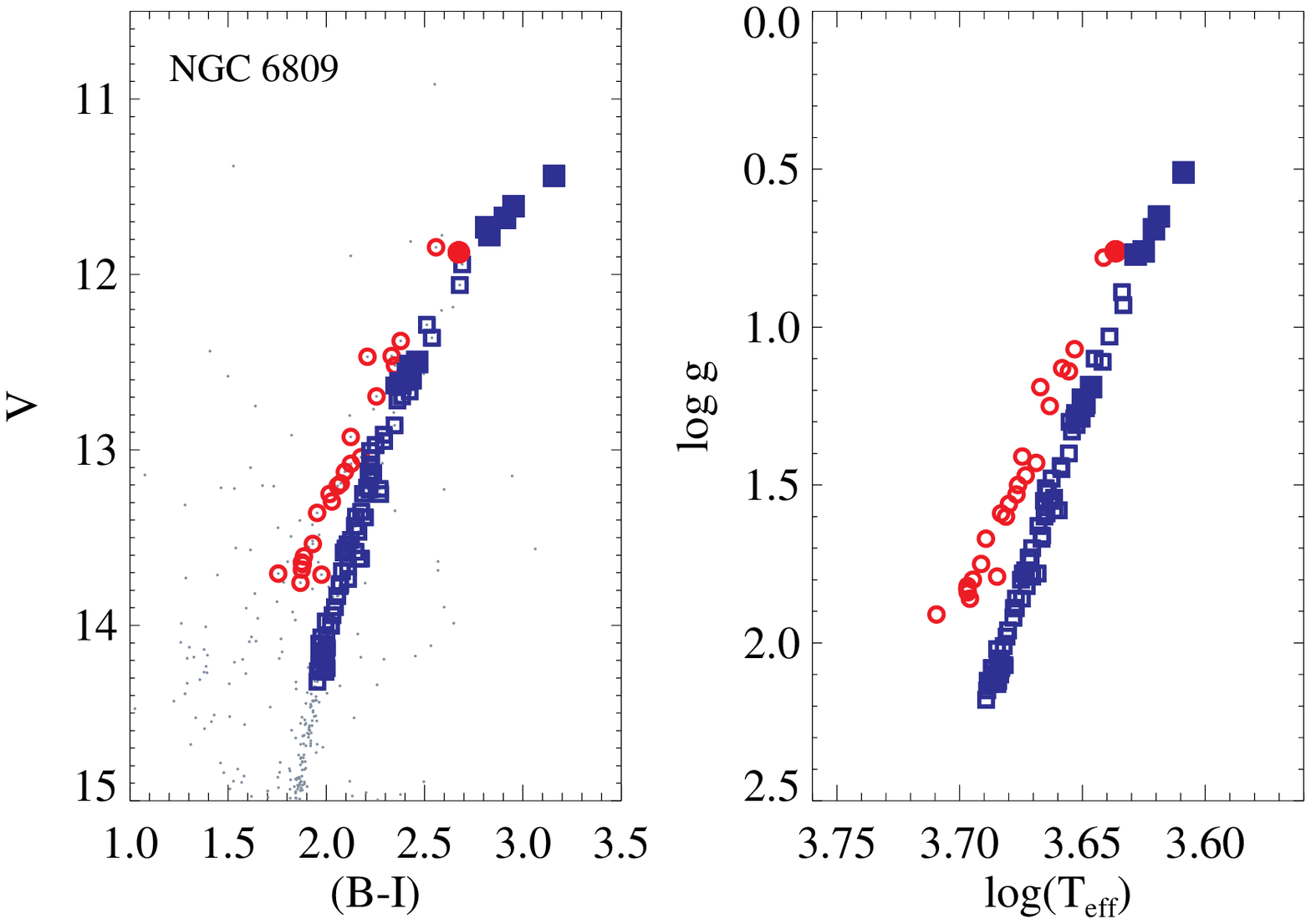}
   \caption{Photometric CMDs (left) and $\log g - \log T_{\rm eff}$ distributions (right) of the cluster member stars. 
            Red circles and blue squares represent AGB and RGB stars, respectively, 
            while the GIRAFFE and UVES samples can be distinguished by open and filled symbols, respectively. 
            The same symbols are used throughout the paper.}
   \label{CMD}
  \end{figure}

As already mentioned in Paper\,I, we selected our targets in NGC\,104, NGC\,6121, and NGC\,6809 from the Johnson-Morgan photometric database 
which is part of the project described in \citet{Stetson2000,Stetson2005} and covers a magnitude range of about three magnitudes for each GC. 
To distinguish AGB from RGB stars, we used several CMDs with different combinations of color indices and magnitudes. We found that in the 
CMDs of, for example, (U-I)$-$U, (U-I)$-$I, and (B-I)$-$V, AGB and RGB stars can be separated efficiently, similarly to \citet{GarciaHernandez2015}. 
Figure \ref{CMD} shows the location of the member stars in the CMDs of (B-I)$-$V.  

All our spectra were obtained with the high-resolution multi-object spectrograph FLAMES, mounted on ESO/VLT-UT2 \citep{Pasquini2003}, taking 
advantage of GIRAFFE (HR 13, HR 15, and HR 19) for the majority of our sample stars and used the UVES fibres (Red 580) for the brightest 
objects of each cluster. Table \ref{obslog} summarizes the most relevant details of our observational campaign. 
Data reduction followed standard procedures and was carried out as described in Paper\,I. The final co-added spectra have signal-to-noise ratios (S/N) 
ranging between 50 and 400 for the GIRAFFE spectra and 50 and 230 for the UVES spectra, depending on the magnitude of the star. We identified 
non-cluster-member stars based on the derived stellar radial velocities and removed them from further analysis. 
The number of stars (initially observed and later confirmed as cluster members) are listed in Table \ref{basic1} for each of the three GCs, together 
with their derived mean barycentric radial velocities, whereas Table \ref{basic2} lists the most relevant information of only the member stars, that is, 
their evolutionary phases (AGB/RGB), the instrument used for collecting the spectrum (GIRAFFE/UVES), and their coordinates, photometry and barycentric 
radial velocities.

\begin{sidewaystable*} \scriptsize
\caption{Basic information of our cluster member stars: evolutionary phase, instrument used for observation, coordinates, photometry and barycentric radial velocity. 
         The complete table is available electronically; we show here the first line of data for each of the three GCs as a guide.}     
\label{basic2}
\centering
\begin{tabular}{c c c c c c c c c c c c c c c c c c c c c}
\hline\hline    
  NGC  & Star ID$^{a}$ & Evol. Ph. & Instrument & RA (J2000) & Dec (J2000) &  $B$  &  $e_{B}$  &  $V$   &  $e_{V}$  &  $I$  &  $e_{I}$  & $J_{2MASS}$ & $e_{J}$ & $H_{2MASS}$ & $e_{H}$ & $K_{2MASS}$ & $e_{K}$ & RV($\mathrm{km\,s}^{-1}$)  \\  
\hline
 104  & AGB58283   & AGB &  GIRAFFE  &  00 22 54.05  &  -72 05 17.00  &  13.802  &  0.0034  &  12.579  &  0.0023  &  11.355  &  0.0035  &  10.414  &  0.024  &  9.699  &  0.025  &  9.573  &  0.025  &  -15.25   \\
$\dots$ & $\dots$ & $\dots$ & $\dots$ & $\dots$ & $\dots$ & $\dots$ & $\dots$ & $\dots$ & $\dots$ & $\dots$ & $\dots$ & $\dots$ & $\dots$ & $\dots$ & $\dots$ & $\dots$ & $\dots$ & $\dots$  \\  

 6121 & AGB30561   & AGB &  GIRAFFE  &  16 23 16.75  &  -26 34 28.00  &  13.170  &  0.0010  &  11.732  &  0.0011  &  10.093  &  0.0033  &   8.851  &  0.029  &  8.038  &  0.034  &  7.832  &  0.023  &   72.04   \\
$\dots$ & $\dots$ & $\dots$ & $\dots$ & $\dots$ & $\dots$ & $\dots$ & $\dots$ & $\dots$ & $\dots$ & $\dots$ & $\dots$ & $\dots$ & $\dots$ & $\dots$ & $\dots$ & $\dots$ & $\dots$ & $\dots$  \\  

 6809 & AGB246868  & AGB &  GIRAFFE  &  19 39 38.65  &  -30 48 32.90  &  14.582  &  0.0028  &  13.757  &  0.0009  &  12.714  &  0.0015  &  11.965  &  0.023  & 11.455  &  0.026  & 11.359  &  0.026  &  171.88   \\
$\dots$ & $\dots$ & $\dots$ & $\dots$ & $\dots$ & $\dots$ & $\dots$ & $\dots$ & $\dots$ & $\dots$ & $\dots$ & $\dots$ & $\dots$ & $\dots$ & $\dots$ & $\dots$ & $\dots$ & $\dots$ & $\dots$  \\  
\hline     
\end{tabular}
\tablefoot{$^{a}$ The `Star ID' reports the original ID from the photometric catalog, to which we added the suffix AGB/RGB to help in the identification of the evolutionary phase of the object.}
\end{sidewaystable*}

\section{Stellar parameters and abundance analysis of our sample stars} 
\label{section:stellarparametersabundances}

\subsection{Effective temperature and surface gravity}
\label{section:Teff-logg}

Optical B, V, and I magnitudes are available for all our stars from the photometric database mentioned in Sect. \ref{section:obsreduction}. 
By cross-matching the coordinates of our targets with the 2MASS catalog \citep{Skrutskie2006} we have also compiled the corresponding J, H, and K 
infrared magnitudes. 

The foreground reddenings and distance moduli adopted for the three GCs are listed in Table \ref{reddening-distance}. 
We note that for the distance modulus of NGC\,104, \citet{Bono2008} took an average of the values derived from the tip of the red giant branch and the 
RR Lyrae methods. For NGC\,6809, by combining information from photometry (V magnitude and B-V) and period data of 13 RR Lyrae stars \citep{Olech1999}, 
we computed the absolute distance modulus using the dual band metal-dependent Period$-$Wesenheit (PWZ) relation recently derived by \citet{Marconi2015}, 
which is almost metallicity independent in B and V. For NGC\,104 and NGC\,6809, we adopted the \citet{Cardelli1989} relations\footnote{A(B) = 4.145\,E(B-V); 
A(V) = 3.1\,E(B-V); A(I) = 1.485\,E(B-V); A(J) = 0.874\,E(B-V); A(H) = 0.589\,E(B-V); A(K) = 0.353\,E(B-V).}, as in Paper\,I. NGC\,6121, instead, is more 
peculiar because it is located in the Galactic plane behind the Sco-Oph cloud complex, and a non-standard reddening law should be applied (\citealp{Hendricks2012}, 
and references therein). For this GC, we thus adopted the recent reddening law and distance modulus reported by \citet{Hendricks2012} who used a combination 
of broadband near-infrared and optical Johnson-Cousins photometry to study the dust properties in the line of sight to this cluster.

\begin{table} \small
\caption{Foreground reddenings, differential reddenings and distance moduli of sample GCs.}     
\label{reddening-distance}
\centering
\begin{tabular}{c c c c}
\hline\hline     
  GC       &    E(B-V)    & $\delta$E(B-V) &  (m-M)$_V$   \\ 
\hline 
 NGC\,104  &  0.04$^{a}$  &  0.028$^{d}$  &  13.40$^{a}$   \\
 NGC\,6121 &  0.37$^{b}$  &  0.200$^{b}$  &  11.28$^{b}$   \\
 NGC\,6809 &  0.11$^{c}$  &  0.027$^{d}$  &  13.61         \\
\hline    
\end{tabular}
\tablefoot{$^{a}$ \citet{Bono2008}; $^{b}$ \citet{Hendricks2012}; $^{c}$ \citet{Richter1999}; $^{d}$ \citet{Bonatto2013}}
\end{table}

As in Paper\,I, we derived the stellar effective temperatures ($T_\mathrm{eff}$) using the \citet{RamirezMelendez2005} photometric calibrations 
for giants and adopting five de-reddened color indices, that is, $(B-V)_{0}$, $(V-I)_{0}$, $(V-J)_{0}$, $(V-H)_{0}$, and $(V-K)_{0}$. 
The variations among these temperature scales are smaller than or comparable to the error of the mean temperature of the five scales. 
So we took the mean value of the temperatures derived from the five color indices as our final $T_\mathrm{eff}$.
The surface gravities $\log g$ were derived from first principles, that is, by using effective temperatures, bolometric corrections (taken from \citealp{Alonso1999}) 
and stellar masses. For the latter, test runs with our stellar evolution code showed that stellar masses on the RGB have a slight dependence on the 
metallicity and age of the cluster, while these differences become negligible on the AGB. Therefore, we assumed $m=0.61\,M_{\odot}$ for the AGB stars in 
all three clusters, whereas we differentiated the values for the RGB stars with $m_\mathrm{RGB,NGC104}=0.91\,M_{\odot}$, $m_\mathrm{RGB,NGC6121}=0.87\,M_{\odot}$ 
and $m_\mathrm{RGB,NGC6809}=0.81\,M_{\odot}$. 
The right panels of Fig. \ref{CMD} show the $\log T_\mathrm{eff} - \log g$ distributions of the member stars. 

It is worth mentioning that all three GCs suffer from some differential reddening. In the case of NGC\,104 and NGC\,6809 the differential reddenings 
are small and comparable in magnitude to the errors on the derived reddenings (cf. Table \ref{reddening-distance} and references therein). If they are 
taken into account as one extra source of uncertainty, the typical errors on their final effective temperatures and gravities become of the order of 
$\pm 70\,\mathrm{K}$ (NGC\,104), $\pm 80\,\mathrm{K}$ (NGC\,6809) and $\pm 0.06$ (for both clusters), respectively. 
The case of NGC\,6121 is however more complex because of its much larger differential reddening ($\sim 0.20\,mag$, \citealp{Hendricks2012}) . 
The errors on $T_\mathrm{eff}$ and $\log g$ could reach $\Delta T_\mathrm{eff} \sim \pm 260\,\mathrm{K}$ and $\Delta \log g \sim \pm 0.12\,\mathrm{dex}$ 
if the reported differential reddening is taken into account at face value, a significant difference from 
$\Delta T_\mathrm{eff} \sim \pm 50\,\mathrm{K}$ and $\Delta \log g \sim \pm 0.03\,\mathrm{dex}$ derived by accounting for only the intrinsic error 
on the reddening. 
However, a range of differential reddening values has been proposed for this cluster, for example, from $\sim 0.05\,mag$ by \citet{Cudworth1990} 
and $\sim 0.10\,mag$ by \citet{Monelli2013,Lardo2017} up to $\sim 0.25\,mag$ by \citet{Mucciarelli2011}. 
Considering the complexity and uncertainties in the reddening of NGC\,6121, we empirically took the median and decided to consider 
$\Delta T_\mathrm{eff} \sim \pm 150\,\mathrm{K}$ and $\Delta \log g \sim \pm 0.08\,\mathrm{dex}$ as representative of our analytical uncertainties.

\subsection{Metallicity and microturbulent velocity}
\label{section:Metallicity}

Metallicity ([Fe/H]) and microturbulence ($\xi_{\rm t}$) were determined as in Paper\,I, where a detailed description of our methodology is provided. In short, metallicities 
were derived by measuring the equivalent widths (EWs) of both \ion{Fe}{i} and \ion{Fe}{ii} unblended lines, restricting our selection to lines with 
EWs between $20\,\mathrm{m}$\AA\,and $120\,\mathrm{m}$\AA. For the computation of the abundances, we used 1D LTE spherical MARCS model atmospheres 
\citep{Gustafsson2008}, the LTE stellar line analysis programme MOOG (\citealp{Sneden1973}, 2014 version) and we assumed a solar iron abundance of 
$\log\epsilon\mathrm{(Fe)}_{\odot}=7.50$ \citep{Asplund2009}. 
Because of their known inter-dependencies, all stellar parameters ($T_\mathrm{eff}$, $\log g$, [Fe/H], $\xi_\mathrm{t}$) were derived iteratively 
and following standard procedures. 
Since standard LTE analyses of \ion{Fe}{i} lines tend to underestimate the true iron abundance, we applied non-LTE (NLTE) corrections to all our 
LTE \ion{Fe}{i} values (\citealp{Lind2012}, and references therein).

Our final stellar parameters are summarized in Table \ref{stellarpar}, while Table \ref{metal} lists the mean metallicities of the AGB and RGB samples 
in the three GCs. For convenience, we have added to this table also the values derived in Paper\,I for NGC\,2808. We note that our RGB results agree 
well with those derived by \citet{Carretta2009a} within the associated errors (cf. two rightmost columns of Table \ref{metal}). A more detailed comparison 
with the literature forms part of Sect.\,4.

Finally, as our overall metallicity of each GC, we chose to use the average value of  [\ion{Fe}{i}/H]$_\mathrm{NLTE}$ and [\ion{Fe}{ii}/H], 
that is, [Fe/H]$_\mathrm{NGC\,104} = -0.82$\,dex, [Fe/H]$_\mathrm{NGC\,6121} = -1.14$\,dex, and [Fe/H]$_\mathrm{NGC\,6809} = -1.86$\,dex.

\begin{table*}  \small
\caption{Stellar parameters of our sample stars. 
         The complete table is available electronically; we show here the first line of data for each GC as a guide.}    
\label{stellarpar}      
\centering       
\begin{tabular}{c c c c c c c c c c c c}    
\hline\hline       
  NGC  &  Star ID  & Evol. Ph. & $T_\mathrm{eff}$ & $\sigma T_\mathrm{eff}~^{a}$ & $\log g$ & $\xi_{\rm t}$ & [\ion{Fe}{i}/H]$_\mathrm{LTE}$ & rms\_lines & [\ion{Fe}{ii}/H] & rms\_lines & [\ion{Fe}{i}/H]$_\mathrm{NLTE}$   \\ 
       &           &           &  ($\mathrm{K}$)  & ($\mathrm{K}$) &  & ($\mathrm{km\,s}^{-1}$) & ($\mathrm{dex}$) & ($\mathrm{dex}$) & ($\mathrm{dex}$) & ($\mathrm{dex}$) & ($\mathrm{dex}$)    \\ 
\hline     
  104  &  AGB58283   &  AGB  &  4291  &  39.3  &   1.23  &   1.60  &  -0.85  &   0.12  &  -0.89  &   0.15  &  -0.82   \\
$\dots$ & $\dots$ & $\dots$ & $\dots$ & $\dots$ & $\dots$ & $\dots$ & $\dots$ & $\dots$ & $\dots$ & $\dots$ & $\dots$  \\ 

  6121 &  AGB30561   &  AGB  &  4433  &  25.3  &   1.28  &   1.75  &  -1.29  &   0.10  &  -1.26  &   0.02  &  -1.24   \\
$\dots$ & $\dots$ & $\dots$ & $\dots$ & $\dots$ & $\dots$ & $\dots$ & $\dots$ & $\dots$ & $\dots$ & $\dots$ & $\dots$  \\ 

  6809 &  AGB246868  &  AGB  &  4964  &  25.7  &   1.86  &   1.16  &  -2.00  &   0.15  &  -1.89  &   0.01  &  -1.91   \\
$\dots$ & $\dots$ & $\dots$ & $\dots$ & $\dots$ & $\dots$ & $\dots$ & $\dots$ & $\dots$ & $\dots$ & $\dots$ & $\dots$  \\ 
\hline       
\end{tabular}
\tablefoot{$^{a}$ $\sigma T_\mathrm{eff}$ is the scatter of the temperatures derived from the five colors we considered.}
\end{table*}

\begin{table*} \small
\caption{Iron abundances of AGB and RGB samples of GCs.}     
\label{metal}
\centering
\begin{tabular}{c c c c c c c c}
\hline\hline      
  GC/sample & [\ion{Fe}{i}/H]$_\mathrm{LTE}$ & [\ion{Fe}{i}/H]$_\mathrm{NLTE}$ & rms([\ion{Fe}{i}/H]$_\mathrm{NLTE}$) & [\ion{Fe}{ii}/H] & rms([\ion{Fe}{ii}/H]) & [\ion{Fe}{i}/H]$_\mathrm{LTE}$ (Carretta) & [\ion{Fe}{ii}/H] (Carretta)  \\ 
            &    $\mathrm{dex}$       &    $\mathrm{dex}$        &     $\mathrm{dex}$        &  $\mathrm{dex}$  &   $\mathrm{dex}$      &     $\mathrm{dex}$            &  $\mathrm{dex}$         \\ 
\hline
 NGC\,104   &  &  &  &  &  &  &    \\
 AGB sample &   -0.82     &   -0.79     &    0.07     &   -0.91    &    0.13     &  &    \\
 RGB sample &   -0.77     &   -0.75     &    0.08     &   -0.82    &    0.15     &  -0.743$\pm$0.047   & -0.769$\pm$0.081  \\
\hline
 NGC\,6121  &  &  &  &  &  &  &    \\
 AGB sample &   -1.21     &   -1.15     &    0.08     &   -1.22    &    0.09     &  &    \\
 RGB sample &   -1.15     &   -1.11     &    0.08     &   -1.15    &    0.06     &  -1.200$\pm$0.043   & -1.197$\pm$0.082  \\
\hline
 NGC\,6809  &  &  &  &  &  &  &    \\
 AGB sample &   -2.03     &   -1.93     &    0.07     &   -1.91    &    0.04     &  &    \\
 RGB sample &   -1.92     &   -1.85     &    0.06     &   -1.84    &    0.06     &  -1.967$\pm$0.041   & -1.933$\pm$0.093  \\
\hline
 NGC\,2808  &  &  &  &  &  &  &    \\
 AGB sample &   -1.19     &   -1.14     &    0.09     &   -1.14    &    0.10     &  &    \\
 RGB sample &   -1.12     &   -1.08     &    0.07     &   -1.09    &    0.07     &  -1.100$\pm$0.059   & -1.160$\pm$0.089  \\ 
\hline     
\end{tabular}
\tablefoot{For convenience, the last rows of the table refer to our results on NGC\,2808 from Paper\,I. 
           The right-most two columns list the metallicities derived from RGB stars by \citet{Carretta2009a} for NGC\,104, NGC\,6121, and NGC\,6809, 
           with the associated errors being the total star-to-star errors in their Table.A.3, while the data for NGC\,2808 are from \citet{Carretta2006}.}
\end{table*}

\subsection{Sodium abundance}
\label{section:Na}

Our stellar Na abundances were derived via spectrum synthesis of the Na doublet at 6154$-$6160\,\AA, using MOOG and MARCS spherical model atmospheres 
interpolated to match our derived stellar parameters. A solar sodium abundance of $\log\epsilon(\mathrm{Na})_{\odot}=6.24$ \citep{Asplund2009} was 
adopted throughout the analysis. As already done for NGC\,2808 (cf. Paper\,I) we took the average of the abundances derived from both doublet lines 
as our final Na abundance. Similarly to iron, Na abundances determined from neutral lines (as in our case) are also affected by the NLTE effect and 
were therefore corrected accordingly. 

We list the Na abundances derived for the individual stars in Table \ref{Na_star}, along with the NLTE-corrected values based on the grids computed 
by \citet{Lind2011}. Table \ref{Na_GC} is similar to Table \ref{metal}, now summarizing the average Na abundances of our three GCs. For convenience, 
it also reports the results from our previous analysis on NGC\,2808.

\begin{table*}  \small
\caption{Na abundances of our sample stars. 
         The complete table is available electronically; we show here the first line of data for each GC as a guide.}    
\label{Na_star}      
\centering      
\begin{tabular}{c c c c c c}     
\hline\hline      
  NGC  &  Star ID  & Evol. Ph. & [Na/H]$_\mathrm{LTE}$ & [Na/H]$_\mathrm{NLTE}$ & [Na/\ion{Fe}{i}]$_\mathrm{NLTE}$   \\ 
       &           &           &   $\mathrm{dex}$      &      $\mathrm{dex}$    &         $\mathrm{dex}$             \\ 
\hline            
  104  &  AGB58283   & AGB &  -0.10  &  -0.21  &   0.61   \\  
$\dots$ & $\dots$ & $\dots$ & $\dots$ & $\dots$ & $\dots$      \\

  6121 &  AGB30561   & AGB &  -0.79  &  -0.85  &   0.39   \\
$\dots$ & $\dots$ & $\dots$ & $\dots$ & $\dots$ & $\dots$      \\

  6809 &  AGB246868  & AGB &  -1.45  &  -1.53  &   0.38   \\
$\dots$ & $\dots$ & $\dots$ & $\dots$ & $\dots$ & $\dots$      \\
\hline         
\end{tabular}
\end{table*}

\begin{table*}  \small
\caption{Mean Na abundances of AGB and RGB samples in our three new GCs}     
\label{Na_GC}      
\centering     
\begin{tabular}{c c c c c c c c}      
\hline\hline      
  GC         & [Na/H]$_\mathrm{LTE}$ & $\delta$NLTE & [Na/H]$_\mathrm{NLTE}$ & $\sigma_\mathrm{[Na/H]}$ & [Na/\ion{Fe}{i}]$_\mathrm{NLTE}$ & $\sigma_\mathrm{[Na/\ion{Fe}{i}]}$ & IQR$_\mathrm{[Na/H]}$    \\ 
             &    $\mathrm{dex}$   & $\mathrm{dex}$ &     $\mathrm{dex}$     &    $\mathrm{dex}$        &    $\mathrm{dex}$                &      $\mathrm{dex}$                &   $\mathrm{dex}$         \\ 
\hline            
 NGC\,104    & & & & & & &   \\
 AGB sample  &  -0.42  &  -0.07  &  -0.49  &  0.19  &  0.30  &  0.20  &  0.285    \\ 
 RGB sample  &  -0.30  &  -0.09  &  -0.39  &  0.17  &  0.36  &  0.19  &  0.330    \\ 
\hline            
 NGC\,6121   & & & & & & &   \\
 AGB sample  &  -0.90  &  -0.07  &  -0.97  &  0.17  &  0.19  &  0.16  &  0.300    \\ 
 RGB sample  &  -0.75  &  -0.08  &  -0.82  &  0.20  &  0.29  &  0.19  &  0.250    \\ 
\hline           
 NGC\,6809   & & & & & & &   \\
 AGB sample  &  -1.60  &  -0.07  &  -1.66  &  0.13  &  0.27  &  0.15  &  0.220    \\ 
 RGB sample  &  -1.56  &  -0.07  &  -1.62  &  0.17  &  0.23  &  0.16  &  0.250    \\ 
\hline
 NGC\,2808   & & & & & & &   \\
 AGB sample  &  -1.00  &  -0.06  &  -1.06  &  0.13  &  0.09  &  0.16  &  0.160    \\ 
 RGB sample  &  -0.98  &  -0.06  &  -1.04  &  0.21  &  0.05  &  0.20  &  0.390    \\ 
\hline      
\end{tabular}
\tablefoot{ {\it $\delta$NLTE} is the NLTE correction of the Na abundance. 
            {\it $\sigma$} represents the dispersion (standard deviation) of the corresponding Na abundance. 
            {\it IQR$_\mathrm{[Na/H]}$} is the inter-quartile range of [Na/H]$_\mathrm{NLTE}$. 
            For convenience, the last section of the table reports our results on NGC\,2808 from Paper\,I.}
\end{table*}

We note that in the two most metal-rich GCs in our sample, the Na doublet was saturated or approaching saturation in 27 stars of NGC\,104 and 
4 stars of NGC\,6121, respectively. These stars were dropped from any further discussion.

\subsection{Error analysis}
\label{section:Error}

We estimated the uncertainties in our derived abundances, following the procedures described in Paper\,I and considering errors of both random 
and systematic nature. 

As random measurement uncertainty, we considered $\sigma/\sqrt{\mathrm{N}}$, where $\sigma$ is the line-to-line dispersion and $\mathrm{N}$ is 
the number of lines measured. A correction according to a t-distribution was applied to the Na and \ion{Fe}{ii} abundances considering the limited 
number of lines present in our spectra (cf. Paper\,I for more details).

For the systematic uncertainty, we selected a total of six stars per GC: 
Four stars observed with GIRAFFE $-$ one cool/hot in each AGB/RGB sample; and two stars observed with UVES-fibre $-$ one cool AGB and one cool RGB, as this 
sub-sample includes only cool stars. With these 18 stars, we tested the effect of varying stellar parameters and EWs (or other key parameters of 
the analysis) by their associated errors on the derived abundances.

Typical total uncertainties, combining both random and systematic sources of error, are summarized in Table \ref{Error} for the GIRAFFE and UVES 
samples of each GC.

\begin{table}  \small
\caption{Total uncertainties of derived Fe and Na abundances.}      
\label{Error}      
\centering      
\begin{tabular}{c c c c c}      
\hline\hline        
                &    [\ion{Fe}{i}/H]    &    [\ion{Fe}{ii}/H]    &         [Na/H]         &   [Na/\ion{Fe}{i}]      \\
                &     $\mathrm{dex}$    &     $\mathrm{dex}$     &     $\mathrm{dex}$     &    $\mathrm{dex}$       \\ 
\hline            
 NGC\,104    &  &  &  &     \\
 GIRAFFE sample &      $\pm 0.07$       &      $\pm 0.12$        &      $\pm 0.13$        &      $\pm 0.13$         \\
 UVES sample    &      $\pm 0.07$       &      $\pm 0.13$        &         $-$            &         $-$             \\
\hline            
 NGC\,6121   &  &  &  &     \\
 GIRAFFE sample &      $\pm 0.13$       &      $\pm 0.14$        &      $\pm 0.16$        &      $\pm 0.11$         \\
 UVES sample    &      $\pm 0.09$       &      $\pm 0.18$        &      $\pm 0.16$        &      $\pm 0.10$         \\
\hline           
 NGC\,6809   &  &  &  &     \\
 GIRAFFE sample &      $\pm 0.11$       &      $\pm 0.07$        &      $\pm 0.13$        &      $\pm 0.12$         \\
 UVES sample    &      $\pm 0.09$       &      $\pm 0.05$        &      $\pm 0.13$        &      $\pm 0.10$         \\
\hline      
\end{tabular}
\end{table}

\subsection{Stars observed with both GIRAFFE and UVES-fibre}

In the case of NGC\,6809, we were able to optimize our fibre configurations by swapping targets between fibre types. We thus observed seven objects 
(one AGB and six RGB) with both GIRAFFE-Medusa and UVES-fibre, which allowed us to test whether or not any zero-point abundance difference exists between the two 
different sets of spectra. We found the following average differences (UVES-fibre minus GIRAFFE): 
$\Delta$[\ion{Fe}{i}/H]$_\mathrm{NLTE}\,=\,+0.02\pm0.04\,\mathrm{dex}$, $\Delta$[\ion{Fe}{ii}/H]$\,=\,-0.03\pm0.07\,\mathrm{dex}$, 
$\Delta$[Na/H]$_\mathrm{NLTE}\,=\,-0.01\pm0.05\,\mathrm{dex}$, and $\Delta$[Na/Fe]$_\mathrm{NLTE}\,=\,-0.03\pm0.06\,\mathrm{dex}$. 
These results show that, despite a lower resolution and smaller spectral coverage, the majority of our spectra (indeed observed with GIRAFFE) match 
very well the results obtained from the analysis of the UVES-fibre counterparts. 
We highlight that in the following Sections, although it does not make any significant difference, we have considered the UVES-fibre results for these 
seven stars.

\subsection{Observed Na abundance distribution along the RGB and AGB} 
\label{section:abund_results}

  \begin{figure}[ht]
    \includegraphics[width=0.46\textwidth]{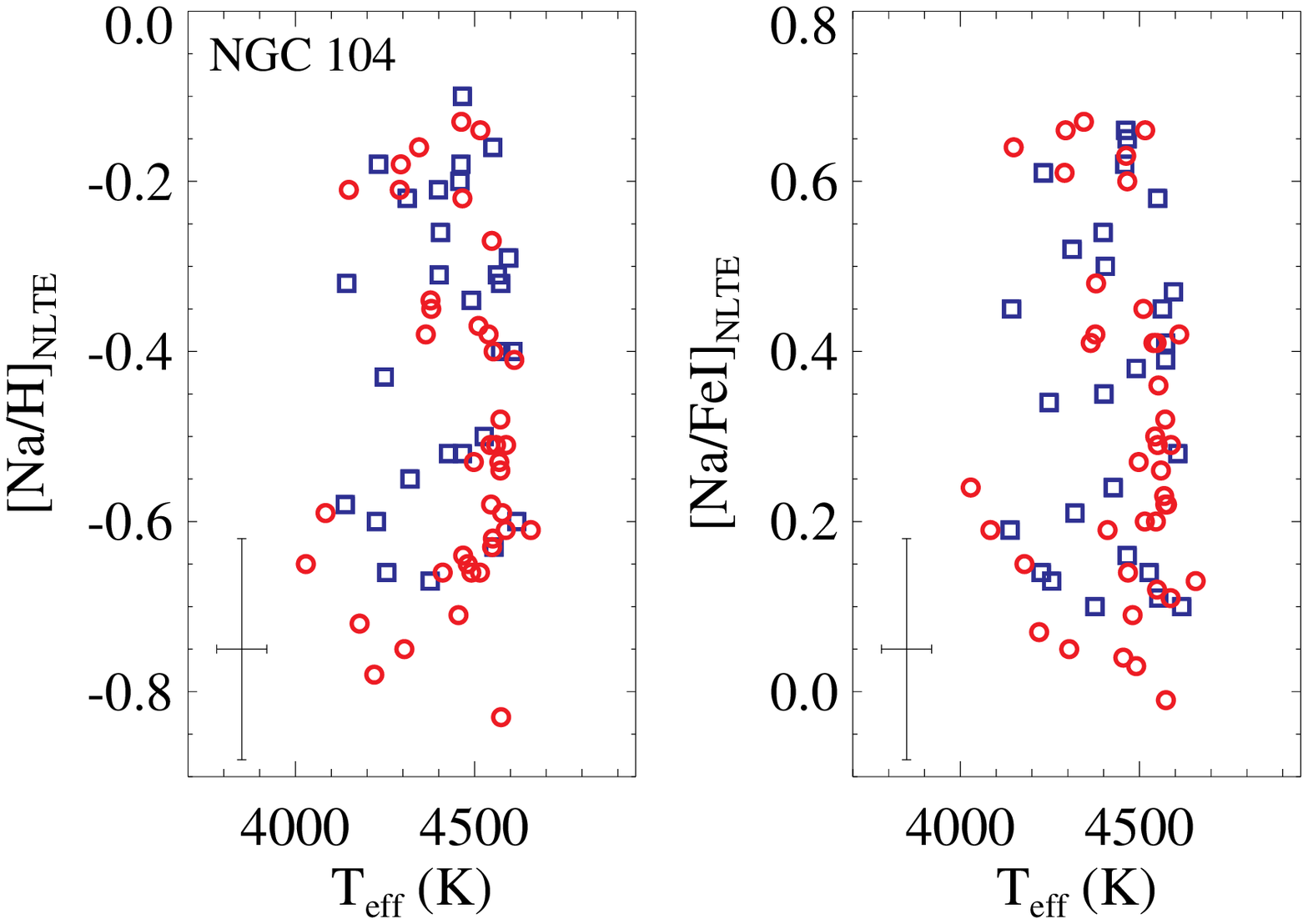}
    \includegraphics[width=0.46\textwidth]{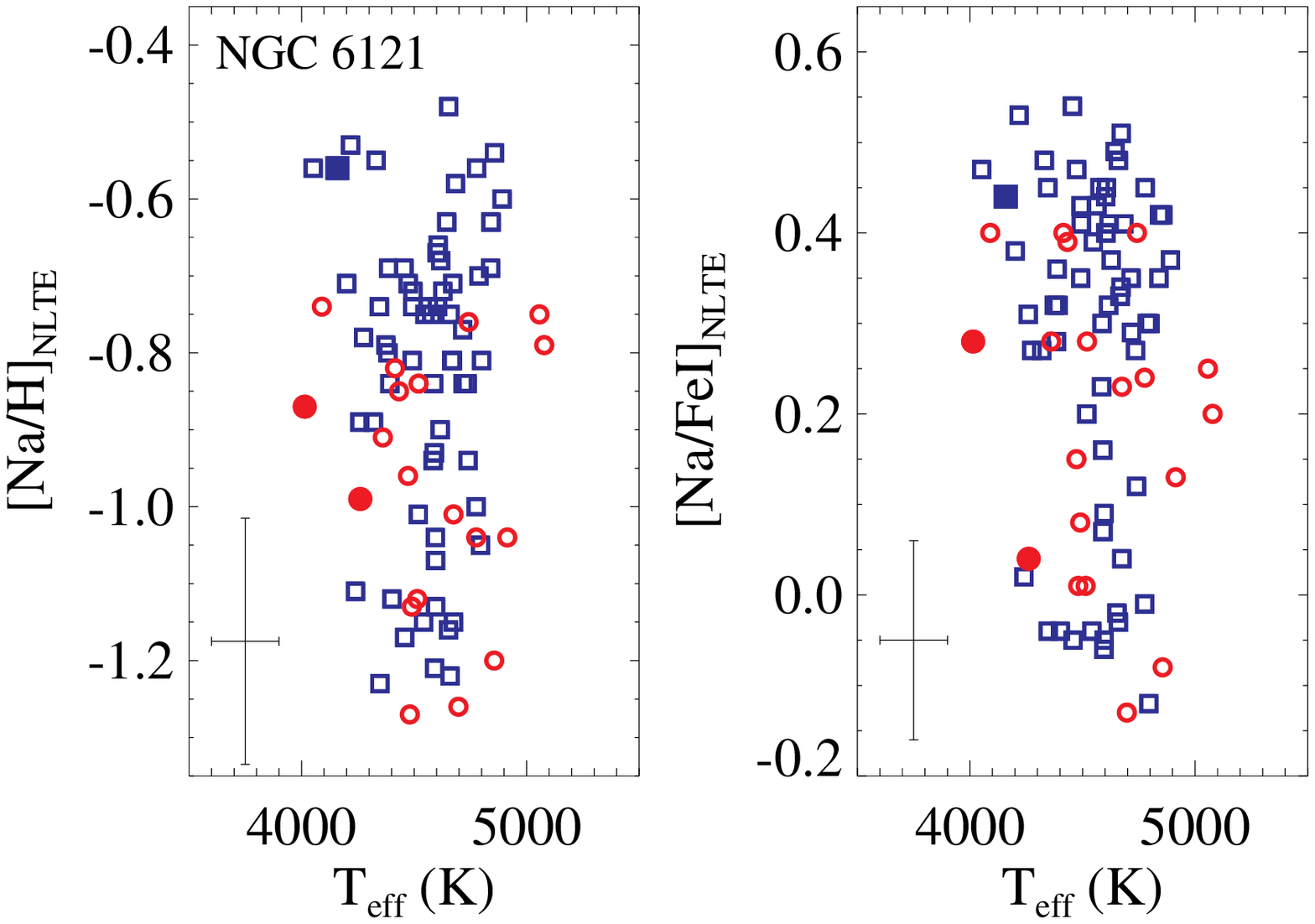}
    \includegraphics[width=0.46\textwidth]{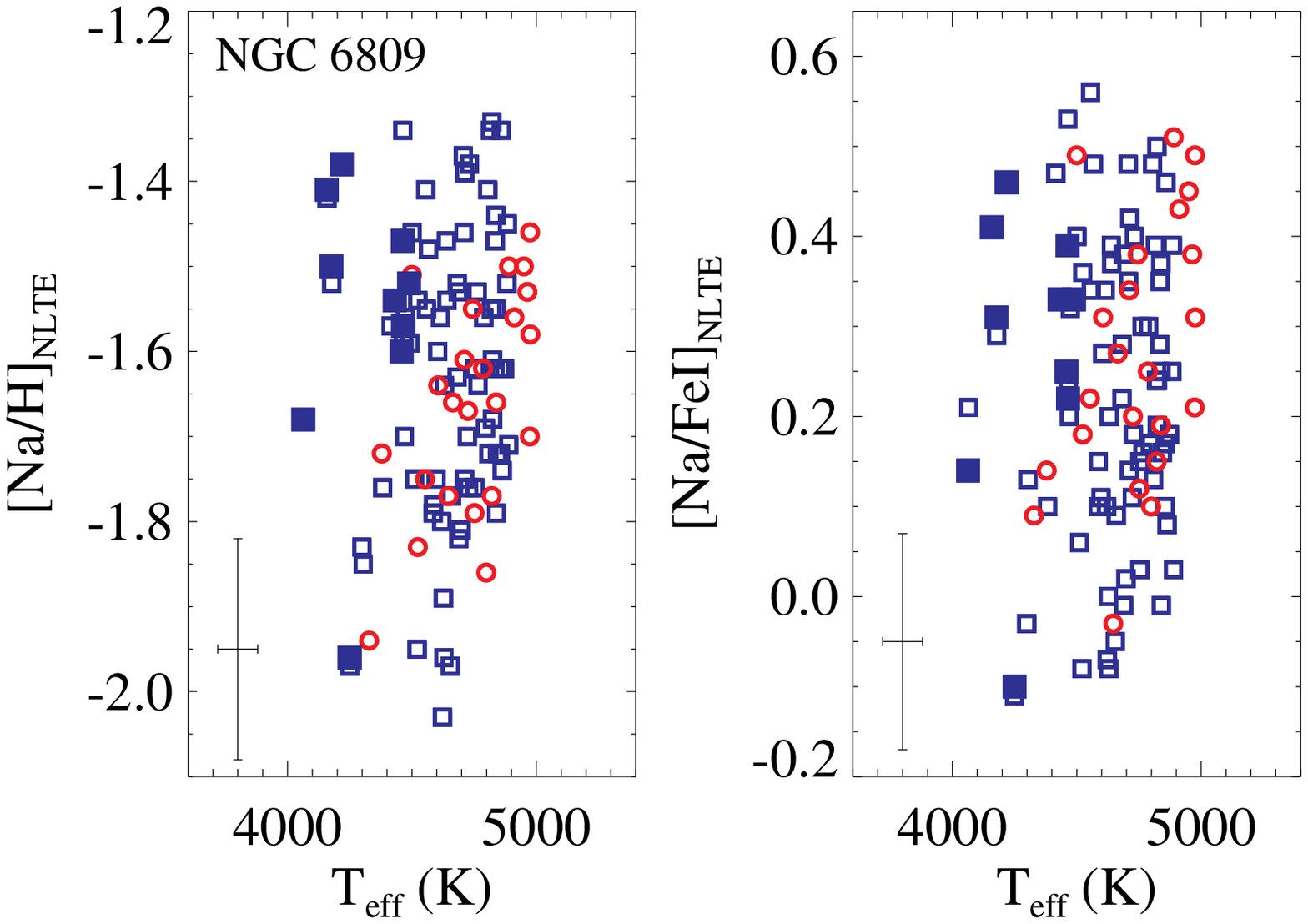}
    \caption{Abundance distributions of our complete (AGB $+$ RGB) sample in NGC\,104, NGC\,6121, and NGC\,6809 (top, middle, and bottom) 
             {\it Left:} [Na/H]$_\mathrm{NLTE}-T_\mathrm{eff}$; 
             {\it right:} [Na/\ion{Fe}{i}]$_\mathrm{NLTE}-T_\mathrm{eff}$.
             Symbols are the same as in Fig. \ref{CMD}. The error bars correspond to our estimates for the GIRAFFE sample.
             }
   \label{Na_distr}
  \end{figure}

Our final Na abundance distributions are shown in Fig.~\ref{Na_distr} as a function of effective temperature for AGB and RGB stars in NGC\,104, NGC\,6121, 
and NGC\,6809. Here, we present them in the form of NLTE [Na/H] and [Na/Fe] (where Fe now only refers to \ion{Fe}{i}) ratios, as listed in Table~\ref{Na_star}.
However, as pointed out by \citet{Campbell2017}, the [\ion{Fe}{i}/H] can be affected by the $T_\mathrm{eff}$ scale, which further influences the degree 
of \ion{Fe}{i}$-$\ion{Fe}{ii} discrepancy that has been found especially in AGB stars by several works \citep{Ivans2001,Lapenna2014,Lapenna2015,Lapenna2016,Mucciarelli2015} 
and also by us (see also Sect. \ref{section:NGC6752}). 
Any uncertainty in the determination of the iron abundance will affect the accuracy of the [Na/Fe] ratios, while [Na/H] is very robust, as also detailed 
by \citet{Campbell2017} who show that parameter variations between studies (caused by using different methods, tools, input data, etc.) have little 
effect on the derived [Na/H]. This was also our conclusion from Paper\,I, which led us to base our discussion of the Na distribution in the RGB 
and AGB samples  on the [Na/H] abundance indicator only.

For each individual cluster we perform a two-sample Kolmogorov-Smirnov (K-S) test to estimate the similarity of the [Na/H] distributions in the AGB and 
RGB samples. All numbers and conclusions discussed below are roughly confirmed by the dispersions ($\sigma$) and the interquartile range (IQR) values of 
the distributions reported in Table \ref{Na_GC}. For each cluster, we also provide a critical summary of how our derived Na abundances compare to other 
abundance studies of similar data quality.

\subsubsection{NGC\,104}
\label{subsubsection:NGC104}

NGC\,104 (40 AGB, 27 RGB) is the GC in which the AGB and RGB samples are the closest in terms of Na abundance distributions, with the AGB stars spanning an even slightly larger range of Na abundances. The K-S test, with D\,=\,0.294 and p-value\,=\,0.101 derived, indicates that the AGB and RGB samples share the 
same [Na/H] distribution at 95$\%$ significance level. 

Our result for this cluster is in very good agreement with \citet{Johnson2015} who found nearly identical [Na/Fe] dispersion in their AGB sample (35 stars) 
to that in the RGB sample (113 stars) analyzed by \citet{Cordero2014}, following the same methodology of \citet{Johnson2015}. 
We carried out a detailed check for 12 AGB and 13 RGB stars in common with our sample (identified by coordinates cross-matching with angular distance < 0.3'') and found mean differences (in the sense of our result minus theirs) of $-78\pm60$\,K in $T_\mathrm{eff}$, $-0.07\pm0.22$ in $\log g$, 
$-0.25\pm0.17$\,km\,s$^{-1}$ in $\xi_\mathrm{t}$, and $-0.12\pm0.10$\,dex in [Fe/H]. 
We assign these differences mainly to the different methods employed to derive the stellar parameters $T_\mathrm{eff}$ and $\log g$ 
(photometry in the present study, spectroscopy in \citealt{Johnson2015} and \citealt{Cordero2014}). 
However, the negligible difference in [Na/H] ($0.02\pm0.12$\,dex; after having checked that their conclusions for [Na/Fe] hold also for the [Na/H] ratio) 
confirms the agreement found in the Na abundance distribution.

\subsubsection{NGC\,6121}
\label{subsubsection:NGC6121}

We find that in NGC\,6121 (19 AGB, 63 RGB), the AGB stars occupy the bottom two-thirds of the [Na/H] distribution of the RGB ones.
There is an actual difference of 0.26\,dex between the maximum [Na/H] values of the AGB and RGB samples, indicating a lack of very Na-rich AGB stars 
in this cluster. The D\,=\,0.408 and p-value\,=\,0.011 from the K-S test confirms that the two populations do not share the same Na distribution.

Our results for this GC agree with those of \citet{MacLean2016} who studied a sample of 106 RGB and 15 AGB stars in NGC\,6121. In their analysis, 
they derived a difference of 0.4\,dex between the maximum [Na/H] values of their AGB and RGB samples (compared to our value of 0.26\,dex, we have more 
Na-rich AGB stars than \citealt{MacLean2016}). 
Although the authors do not provide the coordinates of their sample stars, we were able to find 29 RGB stars in common, by investigating the overlap 
existing between their sample and \citet{Marino2008} and applying our cross-matching criterion with an angular distance $\leq$ 0.3''. 
The mean differences in the derived stellar parameters and Na abundance for this RGB subsample were found to be (in the sense of our result minus theirs) 
$-142\pm78$\,K in $T_\mathrm{eff}$, $-0.05\pm0.13$ in $\log g$, $-0.04\pm0.23$\,km\,s$^{-1}$ in $\xi_\mathrm{t}$, $-0.01\pm0.09$\,dex in [Fe/H], 
and $0.04\pm0.12$\,dex in [Na/H]. Except for the offset in $T_\mathrm{eff}$, we consider this to be a very good agreement. 
Besides, our derived abundances also agree quite well with the results of \citet{Marino2017} who studied 17 AGB stars in this cluster. 
By comparing 14 common stars, differences of $-0.01\pm0.05$\,dex in [Fe/H] and $0.00\pm0.07$\,dex in [Na/H]$_\mathrm{LTE}$ were found between the 
two studies (our result minus theirs), while the differences in $T_\mathrm{eff}$, $\log g$, and $\xi_\mathrm{t}$ were, respectively, $-69\pm54$\,K, 
$-0.09\pm0.17$, and $-0.22\pm0.11$\,km\,s$^{-1}$.

\subsubsection{NGC\,6809}
\label{subsubsection:NGC6809}

In  NGC\,6809 (23 AGB, 77 RGB), we find that the Na abundances of the RGB sample spread more evenly while the AGB stars tend to be more concentrated. 
The difference between the maximum [Na/H] values reached by each sample amounts to 0.12\,dex (the AGB sample reaching lower values). The difference is 
only half of the one observed in NGC\,6121, and the two-sided K-S test does not provide any strong evidence for the two distributions to be different, 
with D\,=\,0.190 and p-value\,=\,0.498. 

This is the first time that  AGB stars are targeted and analyzed for their Na abundances in this cluster. 
We have however 27 RGB stars in common with the \citet{Carretta2009a} sample, for which we found mean differences of $52\pm27$\,K in $T_\mathrm{eff}$, 
$-0.01\pm0.01$ in $\log g$, $-0.16\pm0.38$\,km\,s$^{-1}$ in $\xi_\mathrm{t}$, $0.01\pm0.05$\,dex in [\ion{Fe}{i}/H], 
$0.10\pm0.08$\,dex in [\ion{Fe}{ii}/H], and $0.00\pm0.19$\,dex in [Na/H]. 
Within the errors, our results are consistent with those derived by \citet{Carretta2009a}.

\subsubsection{NGC\,2808}
\label{subsubsection:NGC2808} 

For completeness, we recall here the main results of Paper\,I: the Na abundances of the 33 AGB and 40 RGB stars we analyzed in NGC\,2808 
can be considered to follow the same distribution (K-S test gives D\,=\,0.268 and p-value\,=\,0.137) although a difference of 0.21\,dex was 
found between the maximum [Na/H] values of the AGB and RGB samples, with the AGB maximum value being lower.

\section{Other clusters}
\label{section:OtherGCs}

\subsection{Re-analysis of C13 data for NGC\,6752}
\label{section:NGC6752}
To enlarge the number of clusters studied self-consistently, we decided to reanalyze the publicly available data of C13 following our analytical methods. 
We adopted the reddening E(B-V)\,=\,0.04\,mag and visual distance modulus (m-M)$_V$\,=\,13.24\,mag from \citet{Gratton2003}, and assumed stellar masses 
of $m_\mathrm{AGB}=0.61\,M_{\odot}$ and $m_\mathrm{RGB}=0.83\,M_{\odot}$, following the theoretical predictions of stellar evolution models \citep[assuming an age of 12.5 Gyr,][]{Chantereau2015}. 
Using the methods described in Sects. \ref{section:Teff-logg} and \ref{section:Metallicity}, we derived the stellar parameters and the metallicities 
for a total of 44 individual stars (20 AGB and 24 RGB stars). Overall, we find a good agreement with the C13 result on the effective temperature with a mean 
difference of $\Delta\,T_\mathrm{eff} = -24\pm51\,\mathrm{K}$ (ours minus C13), while the mean differences on the gravity and microturbulent velocity 
are $\Delta\,\log g = -0.14\pm0.07$ and $\Delta\,\xi_\mathrm{t} = -0.30\pm0.25\,\mathrm{km\,s}^{-1}$, respectively. 
We derive the following LTE Fe abundances: 
[\ion{Fe}{i}/H]$_\mathrm{AGB} = -1.85\pm0.02$\,dex ($\sigma = 0.06$\,dex), 
[\ion{Fe}{ii}/H]$_\mathrm{AGB} = -1.65\pm0.06$\,dex ($\sigma = 0.04$\,dex), 
[\ion{Fe}{i}/H]$_\mathrm{RGB} = -1.66\pm0.02$\,dex ($\sigma = 0.07$\,dex), and 
[\ion{Fe}{ii}/H]$_\mathrm{RGB} = -1.55\pm0.06$\,dex ($\sigma = 0.05$\,dex), and find for the AGB stars a very good agreement with \citet{Lapenna2016}, 
who re-observed the 20 AGB stars of C13 at higher resolution with ESO-VLT/UVES ([\ion{Fe}{i}/H]$_\mathrm{AGB} = -1.80\pm0.01$\,dex and 
[\ion{Fe}{ii}/H]$_\mathrm{AGB} = -1.58\pm0.01$\,dex). 
After applying NLTE corrections to the \ion{Fe}{i} abundances \citep{Bergemann2012,Lind2012}, we derived the average values of 
[\ion{Fe}{i}/H]$_\mathrm{NLTE,AGB} = -1.76\pm0.04$\,dex and [\ion{Fe}{i}/H]$_\mathrm{NLTE,RGB} = -1.60\pm0.04$\,dex, 
which bring the \ion{Fe}{i}$-$\ion{Fe}{ii} values for the RGB stars into agreement (within the errors) and almost halves the \ion{Fe}{i}$-$\ion{Fe}{ii} 
difference for the AGB stars.  
Compared to the value assumed by C13 for all their sample stars ([Fe/H]\,=\,$-$1.54\,dex), our derived metallicities are slightly lower and point to 
a 0.1\,dex difference between AGB and RGB stars.

 \begin{figure}
   \centering
   \includegraphics[width=0.46\textwidth]{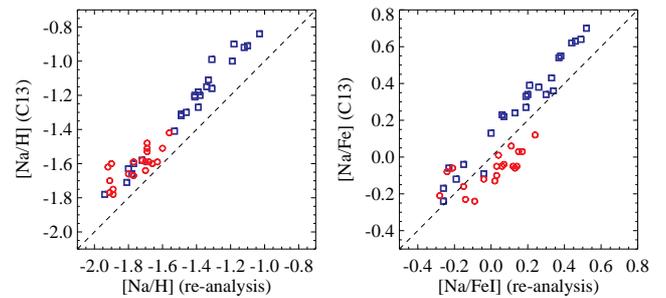}
   \caption{Comparison of the Na abundance (NLTE) of NGC\,6752 sample between our derived results and those of C13. 
            Red circles represent AGB stars, and blue squares represent RGB stars}
   \label{comp_Campbell2013}
  \end{figure}

Considering the weakness of the $6154-60$\,\AA\ Na lines in a non-negligible number of spectra of the C13 sample, we followed their choice and derived 
the Na abundances from the EWs of the Na doublet at $5682-5688$\,\AA \footnote{
The choice of the Na doublet has only a negligible impact on the derived abundances ($-$0.027\,dex, $\sigma$\,=\,0.073\,dex, {\it{bluer $-$ redder}} doublet).}. 
We apply the same NLTE corrections \citep{Lind2011} to the derived Na abundances as for the other GCs and find a mean difference (ours minus C13) 
of $-0.16\pm0.06$\,dex in [Na/H] and of $-0.03\pm0.13$\,dex in [Na/Fe]. 
We compare the derived Na abundances from the two studies in Fig.~\ref{comp_Campbell2013}. 
The differences in [Na/H] appear to be systematic (likely due to the combination of different stellar parameters and adopted solar abundances), 
while the effect on [Na/Fe] is influenced mainly by the adopted Fe abundance (overall cluster metallicity in the case of C13, individual stellar 
metallicities in our case; see also \citealp{Lapenna2016}). The remaining offsets are likely to come from the NLTE correction applied to 
the Na abundance (C13 used the \citealp{Gratton1999} values).

\begin{figure}
   \centering
   \includegraphics[width=0.46\textwidth]{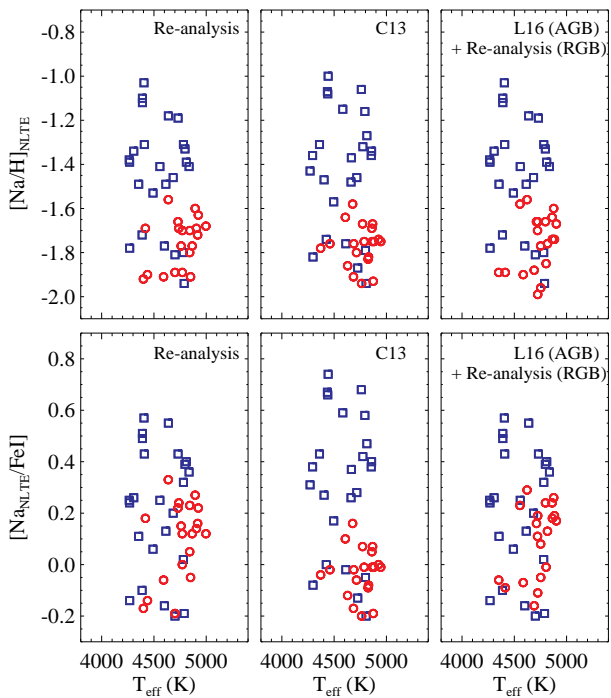}
   \caption{Na abundance (NLTE) distributions of the sample of C13 in NGC\,6752.
            The left-column panels show our re-analysis results, middle-column panels show the Na abundance distribution from C13, 
            and the right-column panels show the Na abundance of the AGB sample derived by \citet{Lapenna2016} together with the RGB stars from our re-analysis. 
            For convenience of comparison, the data from C13 and \citet{Lapenna2016} are shifted systematically to our scale with a constant 
            which is equal to the mean difference between their results and ours, while the dispersions derived in each study are kept. 
            Red circles and blue squares represent AGB and RGB stars, respectively.
            }
   \label{Na_NGC6752}
  \end{figure}

Figure~\ref{Na_NGC6752} shows all results of this re-analysis (left column), with a direct comparison to the values published by C13 (middle column). 
For AGB stars, our [Na/H] distribution agrees quite well with that of C13, while this is not the case for [Na/Fe]. 
Testing the assumption on the metallicities made by the two analyses shows that we can fully reproduce C13 results as soon as we use an overall 
metallicity value for the cluster. 
Both our [Na/H] and [Na/\ion{Fe}{i}] distributions derived for the AGB sample agree well with those derived by \citet{Lapenna2016} from higher-resolution spectral data, as shown in the right column of Fig.~\ref{Na_NGC6752} (labeled as L16; the RGB stars are from our re-analysis and are 
shown only to aid the comparison), which supports our abundance results.
As in Sect. \ref{section:abund_results}, we take [Na/H] as the Na abundance indicator and find that there is a significant lack of Na-rich AGB stars 
in the sample of NGC\,6752 (2P AGB stars account for $\sim$15\%, cf. Sect. \ref{section:discussion}).
However, \citet{Lapenna2016} claimed that both 1P and 2P stars populate the AGB of NGC\,6752 with $\sim$65\% of AGB stars belonging to 2P based on 
the [O/Fe]-[Na/Fe] distribution. Only stars with extreme Na enhancements are claimed to be missing in their AGB sample. 
We note, however, that this conclusion was derived based on [Na/Fe] ratios. 
The \ion{Fe}{i}$-$\ion{Fe}{ii} discrepancy (especially their different behaviors in AGB and RGB stars) may affect the relative Na abundance distributions 
between AGB and RGB stars (e.g., Fig.~\ref{Na_NGC6752}, left column) so that the [Na/Fe] indicator carries a larger uncertainty compared to 
the [Na/H] (Sect. \ref{section:abund_results}). 
Furthermore, we note that our derived fraction of 2P-AGB stars ($\sim$15\%) is lower than the one predicted by \citet[$\sim$50\%]{Cassisi2014b} 
based on simulations of the horizontal branch. While the specific fraction numbers of different stellar populations depend on the adopted separation 
criteria (see also Sect. \ref{section:1Pvs2P}), the sampling could also affect the results. Although we believe that this GC deserves further 
scrutiny, for the sake of a coherent discussion, we use our derived abundance ratios in the following Sections.

\subsection{Literature data for four additional GCs}
\label{section:literature-GCs}

In the literature, one finds four other GCs whose AGB stars have been targeted for their Na abundances through moderate- and high-resolution 
spectroscopic observations: NGC\,5904 \citep{Ivans2001,Lai2011}, NGC\,5986 \citep{Johnson2017}, NGC\,6205 \citep{Johnson2012}, and NGC\,6266 \citep{Lapenna2015}. 
The Na abundance patterns observed in these clusters are shown in Fig.~\ref{NaH_5GCs-literature}, where the data points have already been adjusted 
to be on the same solar abundance and NLTE-correction scales as our own data set; for NGC\,5904 the abundance data from \citet{Lai2011} 
have been unified to the system of \citet{Ivans2001} according to the common star between the two studies, while for the common star we adopt the 
data derived by \citeauthor{Ivans2001}. 
 
{\bf{NGC\,6205}} (M13; [Fe/H] = $-$1.57, \citealt{Johnson2012}) seems to be relatively devoid of Na-normal AGB stars, with a difference of $\sim$\,0.4\,dex 
between the [Na/H] minimum values in the RGB and the AGB samples. 
But, as suggested by \citet{GarciaHernandez2015}, due to the difficulty of distinguishing AGB from RGB stars at the bright end of the giant branch, 
some misclassification might have occurred, which could slightly favor the RGB sample. Notwithstanding, the presence of a large fraction of Na-rich 
AGB stars in NGC\,6205 is clear. 

{\bf{NGC\,5986}} ([Fe/H] = $-$1.54, \citealt{Johnson2017}) shows comparable [Na/H] spreads in AGB and RGB star samples with the maximum [Na/H] 
value of AGB stars being 0.14\,dex lower than that of RGB stars. The sample of AGB stars is, however, rather limited.

{\bf{NGC\,5904}} (M5; [Fe/H] = $-$1.22, \citealt{Ivans2001,Lai2011}) shows a paucity of very Na-rich AGB stars compared to RGB stars, with a 
difference of 0.25\,dex between the maximum [Na/H] value of the AGB and RGB samples.

{\bf{NGC\,6266}} (M62; [Fe/H] = $-$1.05, \citealt{Lapenna2015}) makes itself distinct by showing no Na-rich AGB star in the small sample analyzed 
by \citet{Lapenna2015}. However, in our opinion, no firm conclusion can be drawn because the AGB phase may not be sufficiently sampled.

\section{Discussion for the full sample of GCs with Na abundance determination on the AGB} 
\label{section:discussion}

  \begin{figure*}[ht]
   \centering
   \includegraphics[width=0.94\textwidth]{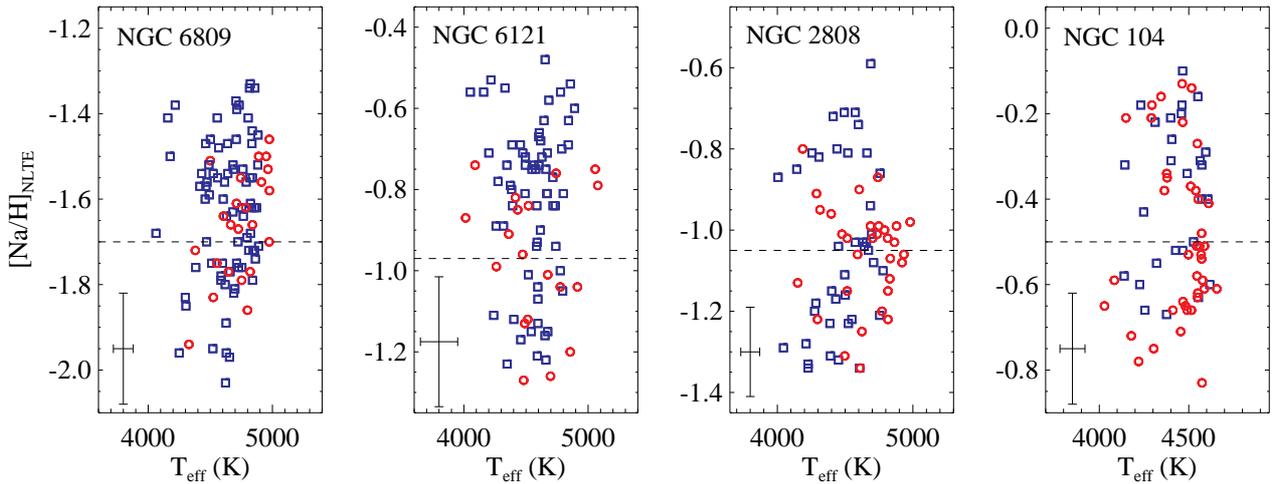}
    \caption{[Na/H] versus $T_\mathrm{eff}$ of the AGB and RGB stars (red and blue symbols, respectively) in  
             NGC\,6809, NGC\,6121, NGC\,2808, and NGC\,104 from the present analysis and Paper\,I. 
             The horizontal black dashed lines mark the critical [Na/H] ratio separating roughly the Na-poor 1P and Na-rich 2P stars according 
             to \citet{Carretta2009a} criteria. 
             The clusters are presented by increasing [Fe/H] from left to right.
             }
  \label{NaH_4GCs}
  \end{figure*}
  
  \begin{figure*}[ht]
   \centering
   \includegraphics[width=0.96\textwidth]{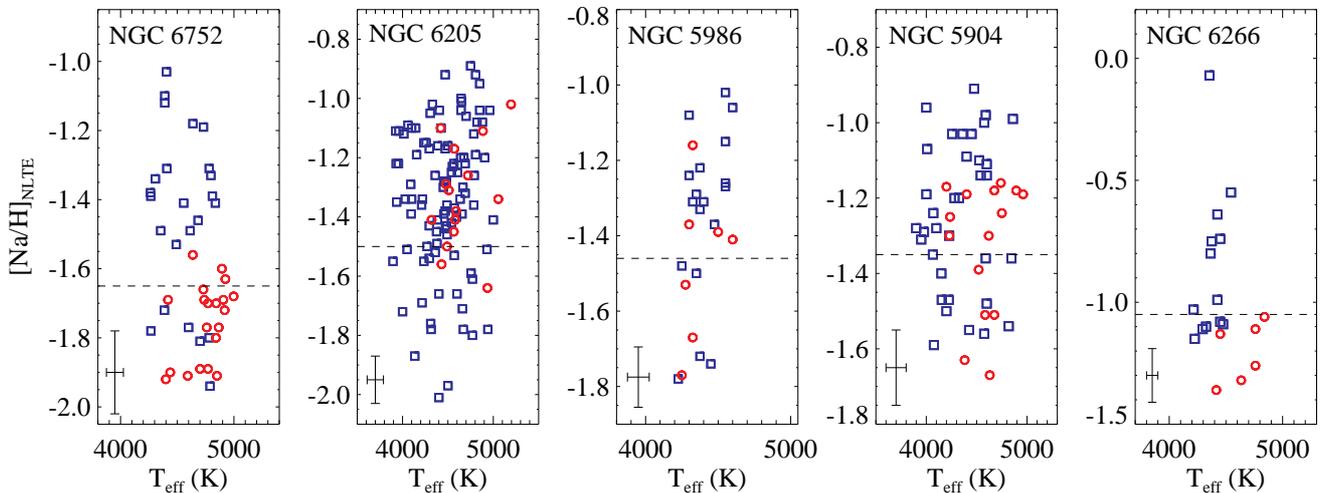}
     \caption{Same as Fig.~\ref{NaH_4GCs} but for NGC\,6752, NGC\,6205, NGC\,5986, NGC\,5904, and NGC\,6266. 
              The values for NGC\,6752 are from our re-analysis of C13 data, 
              the others are from the literature (see the text) but the adopted solar abundance and the NLTE correction for Na 
              have been homogenized to those we used for our own GCs.
              }
     \label{NaH_5GCs-literature}
  \end{figure*}

\subsection{Comparison criteria}
\label{subsection:criteria}

For the reasons previously described, we use the [Na/H]$_\mathrm{NLTE}$ data to discuss the differences and similarities in terms of Na abundances 
in the nine GCs for which both the RGB and AGB have been studied. 
The abundance determination is self-consistent for five GCs (NGC\,6809, NGC\,6121, NGC\,2808, NGC\,104, and NGC\,6752; this paper and Paper I). 
For the four other GCs (NGC\,6205, NGC\,5986, NGC\,5904, and NGC\,6266), we use the data from the original papers that we modified to have 
consistent Na reference solar abundance and NLTE corrections (Sect. \ref{section:literature-GCs}). 

We use the same definition as in Paper\,I to distinguish Na-normal and Na-rich stars (often called 1P and 2P stars in the literature; 
see e.g., \citealp{PrantzosCharbonnel2006,Carretta2009a}), the latter ones being defined as those having [Na/H] higher than 
[Na/H]$_\mathrm{cri}$\,=\,[Na/H]$_\mathrm{min}$\,+\,0.3\,dex, where $ $[Na/H]$_\mathrm{min}$ is the minimum Na value derived for the RGB+AGB 
sample in a given cluster and 0.3\,dex is about one third of the [Na/H] spread.  
Table~\ref{9GCs_comp} gathers the values of [Na/H]$_\mathrm{cri}$, of the fraction of Na-rich RGB and AGB 2P stars 
(f$_\mathrm{2P,RGB}$ and f$_\mathrm{2P,AGB}$ respectively), and of the Na spreads $\Delta$[Na/H] for the RGB and AGB subsamples in each GC; 
we have also collected in the table important cluster characteristics that are relevant for the discussion.

\subsection{Na abundance distributions among the RGB and AGB samples}
\label{section:comparisonall}

\begin{figure}[ht]
   \centering
   \includegraphics[width=0.38\textwidth]{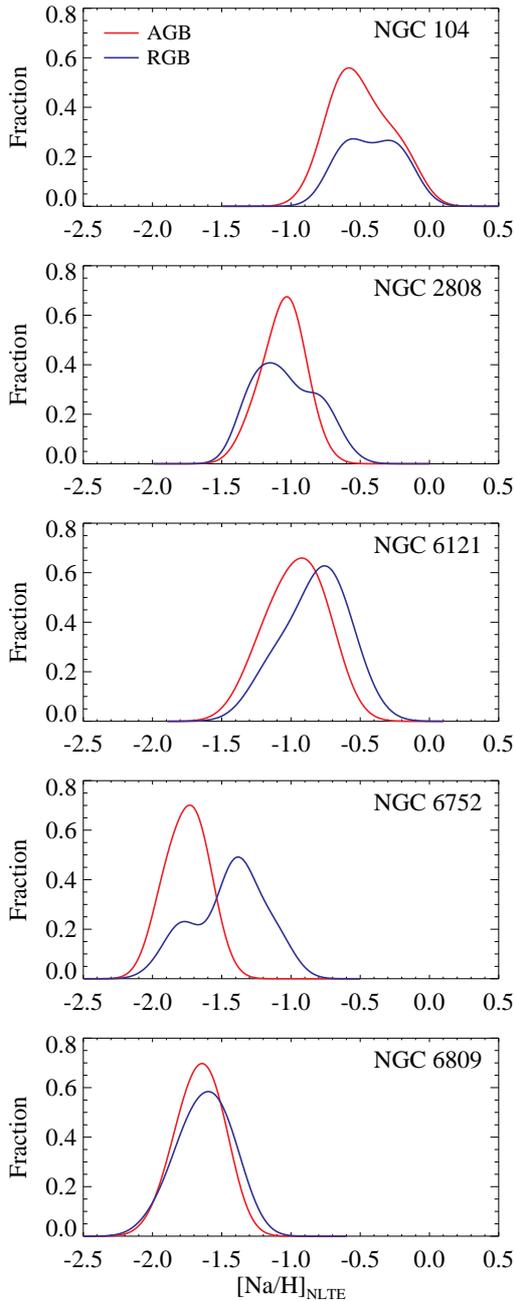}
     \caption{Continuous histograms of [Na/H] of the AGB (red) and RGB (blue) samples for the five GCs we have analyzed 
              in a self-consistent way (whose [Fe/H] decrease from top to bottom).}
  \label{Na_hist_5GCs}
  \end{figure}

\begin{figure}[ht]
   \centering
   \includegraphics[width=0.38\textwidth]{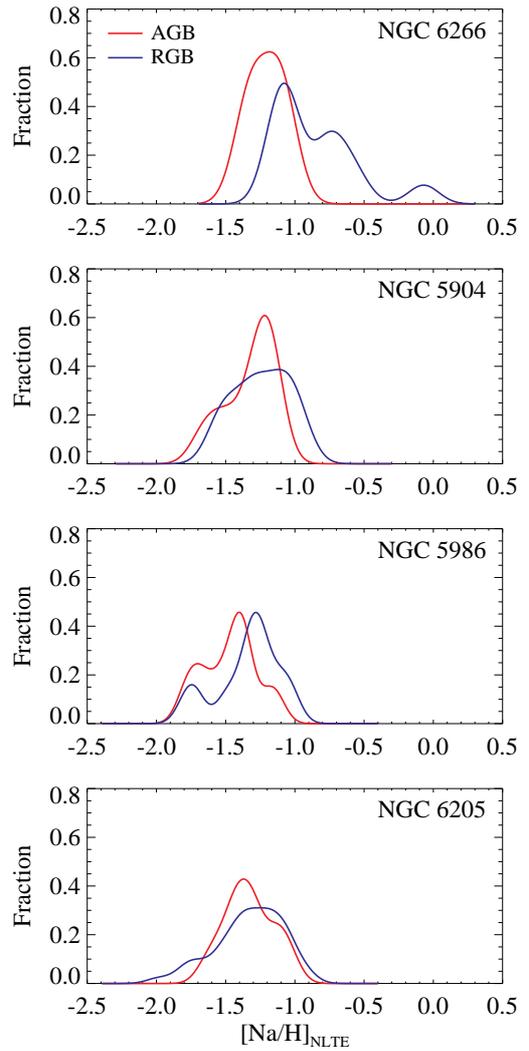}
     \caption{Same as Fig.~\ref{Na_hist_5GCs} but for literature data.}
  \label{Na_hist_4GCs}
  \end{figure}

We gather all the [Na/H]$_\mathrm{NLTE}$ data for the nine GCs as a function of stellar effective temperature (Figs.~\ref{NaH_4GCs} and 
\ref{NaH_5GCs-literature}) and in the form of continuous histograms (Figs.~\ref{Na_hist_5GCs} and \ref{Na_hist_4GCs}, where every star is represented 
by a Gaussian profile with a weight of one and standard deviation equalling the uncertainty on the measurement) for RGB and AGB samples.
In all cases, the Na dispersion observed for RGB and AGB stars does not depend on the effective temperature (Figs.~\ref{NaH_4GCs} and 
\ref{NaH_5GCs-literature}) or the brightness of the stars (not shown here). This means that there is no in situ evolution effect that modifies 
the Na abundance inside the GC evolved stars we observe today.

A quick look at these figures shows that in the majority of the GCs under scrutiny (eight out of nine, the only exception being NGC\,104), 
the Na spread is smaller among AGB stars than among RGB stars (see the actual dispersion numbers in Table~\ref{9GCs_comp}).  
More specifically, the maximum [Na/H] values derived for the AGB stars are lower than the ones derived for the RGB stars. 
In three out of these eight GCs (NGC\,6809, NGC\,6205, and NGC\,5986), the maximum values for Na on the RGB and the AGB are however 
marginally consistent considering the errors.
Three clusters (NGC\,2808, NGC\,5904, NGC\,6121) clearly lack the most Na-rich AGB star. 
Finally, the last two clusters (NGC\,6752 and NGC\,6266\,\footnote{Larger uncertanity may exist for this cluster considering the paucity of stars 
that have been studied so far \citep{Lapenna2015}.}) 
stand out by showing (almost) no Na-rich AGB stars (i.e., with [Na/H] higher than [Na/H]$_\mathrm{cri}$) in the samples analyzed so far. 

\begin{table*} \small
\caption{Na abundance spread, critical value of the Na abundance distinguishing 1P and 2P stars, 
         and the corresponding fraction of Na-rich 2P stars f$_\mathrm{2P}$ (together with the associated errors based on Beta distribution) 
         in the RGB and AGB samples. 
         The five GCs on the left side of the vertical line are the GCs analyzed (or re-analyzed) by ourselves homogeneously, and their [Fe/H] are 
         the values derived by us, while the four GCs on the right are collected from the literature (see the text).
         The other global GC properties are from the literature. 
        } 
\label{9GCs_comp} 
\centering      
\begin{tabular}{c c c c c c | c c c c}      
\hline\hline     
 \noalign{\smallskip}
                               & NGC\,6809   & NGC\,6121   & NGC\,2808   & NGC\,104    & NGC\,6752   & NGC\,6205   & NGC\,5986   & NGC\,5904   & NGC\,6266   \\    
                               &   M\,55     &   M\,4      &             & 47\,Tuc     &             &   M\,13     &             &   M\,5      &   M\,62     \\ 
\hline                                                                                                                                                      
\noalign{\smallskip}                                                                                                                                        
 $ $[Na/H]$_\mathrm{cri}$      &  -1.70      &  -0.97      &  -1.05      &  -0.50      &  -1.65      &  -1.50      &  -1.46      &  -1.35      &  -1.05      \\
\hline                                                                                                                                                      
\noalign{\smallskip}                                                                                                                                        
 $\Delta$[Na/H]$_\mathrm{AGB}$ &  0.48       &  0.53       &  0.54       &  0.70       &  0.36       &  0.62       &  0.61       &  0.51       &  0.30       \\
   f$_\mathrm{2P,AGB}$         &  65$\pm$11  &  53$\pm$11  &  55$\pm$9   &  40$\pm$8   &  15$\pm$11  &  87$\pm$13  &  57$\pm$18  &  67$\pm$13  &  0$\pm$23   \\
\hline                                                                                                                                                      
\noalign{\smallskip}                                                                                                                                        
 $\Delta$[Na/H]$_\mathrm{RGB}$ &  0.70       &  0.75       &  0.75       &  0.57       &  0.91       &  1.12       &  0.76       &  0.68       &  1.08       \\
   f$_\mathrm{2P,RGB}$         &  66$\pm$6   &  76$\pm$6   &  52$\pm$8   &  67$\pm10$  &  75$\pm$11  &  76$\pm$5   &  72$\pm$12  &  69$\pm$9   &  62$\pm$14  \\ 
\hline                                                                                                                                                      
\noalign{\smallskip}                                                                                                                                        
 $ $[Fe/H]                     &  -1.86      &  -1.14      &  -1.11      &  -0.82      &  -1.60      &  -1.57      &  -1.54      &  -1.22      &  -1.05      \\
    Age                        &  13.00      &  11.50      &  11.00      &  11.75      &  12.50      &  12.00      &  12.25      &  11.50      &  11.60      \\ 
                               & $\pm$0.25   & $\pm$0.38   & $\pm$0.38   & $\pm$0.25   & $\pm$0.25   & $\pm$0.38   & $\pm$0.75   & $\pm$0.25   & $\pm$0.60   \\ 
    M$_\mathrm{V}$             &  -7.57      &  -7.19      &  -9.39      &  -9.42      &  -7.73      &  -8.55      &  -8.44      &  -8.81      &  -9.18      \\ 
    mass                       &  0.269      &  0.195      &  1.420      &  1.500      &  0.317      &  0.775      &  0.599      &  0.857      &  1.220      \\ 
    HBR                        &   0.87      &  -0.06      &  -0.49      &  -0.99      &   1.00      &   0.97      &   0.97      &   0.31      &   0.32      \\
    r$_\mathrm{h}$             &   4.46      &   2.34      &   2.12      &   3.65      &   2.72      &   3.34      &   3.18      &   4.60      &   2.47      \\
    r$_\mathrm{t}$             &  25.10      &  20.79      &  43.42      &  56.10      &  64.39      &  56.40      &  31.83      &  61.96      &  18.00      \\
    ellipticity                &   0.02      &   0.00      &   0.12      &   0.09      &   0.04      &   0.11      &   0.06      &   0.14      &   0.01      \\
    c.c.                       &   0.93      &   1.65      &   1.56      &   2.07      &   2.50      &   1.53      &   1.23      &   1.73      &   1.71      \\
    $\sigma_\mathrm{v}$        &   4.0       &   4.0       &  13.4       &  11.0       &   4.9       &   7.1       &   $-$       &   5.5       &  14.3       \\
\hline       
\end{tabular} 
 \tablefoot{The GC ages (Gyr) are adopted from \citet{VandenBerg2013}, except NGC\,6266 whose age is from \citet{Roediger2014}; 
            the M$_\mathrm{V}$ (mag), ellipticities, c.c., and $\sigma_\mathrm{v}$ ($\mathrm{km}\,\mathrm{s}^{-1}$) are from \citet[2010 version]{Harris1996};        
            the masses ($\times10^{6}\,\mathrm{M_{\sun}}$) are adopted from \citet{Boyles2011};          
            the HBR, r$_\mathrm{h}$ ($\mathrm{pc}$), and r$_\mathrm{t}$ ($\mathrm{pc}$) are from \citet{Mackey2005}.}
\end{table*}

The continuous histograms of the [Na/H] (Figs. \ref{Na_hist_5GCs} and \ref{Na_hist_4GCs}) indicate that the Na abundance distributions of 
the RGB and AGB stars cannot be described by one type of profile (single- or double-peak) and confirm that they vary from cluster to cluster. 
According to the current data, 
both AGB and RGB samples of NGC\,104 and NGC\,6205 are bimodal; 
those of NGC\,6121 and NGC\,6809 are unimodal; 
for NGC\,2808 and NGC\,6752, their RGB and AGB samples appear to be bimodal and unimodal, respectively; 
NGC\,6266 has a unimodal AGB sample and trimodal RGB sample; 
NGC\,5904 shows bimodal distribution in AGB sample but a broadened profile in RGB sample which indicates that three closely located peaks may exist; 
while NGC\,5986 shows trimodal distributions in both AGB and RGB samples. 
Separations can clearly be found between the main peaks of AGB and RGB samples of NGC\,6121, NGC\,6752, NGC\,6266, and NGC\,5986,
with the ones of AGB always having lower [Na/H] values than RGB, and the main peaks in NGC\,6752 separate the most.

\subsection{Fractions of 1P and 2P stars}
\label{section:1Pvs2P}

\begin{figure*}[ht]
   \centering
    \includegraphics[width=0.47\textwidth]{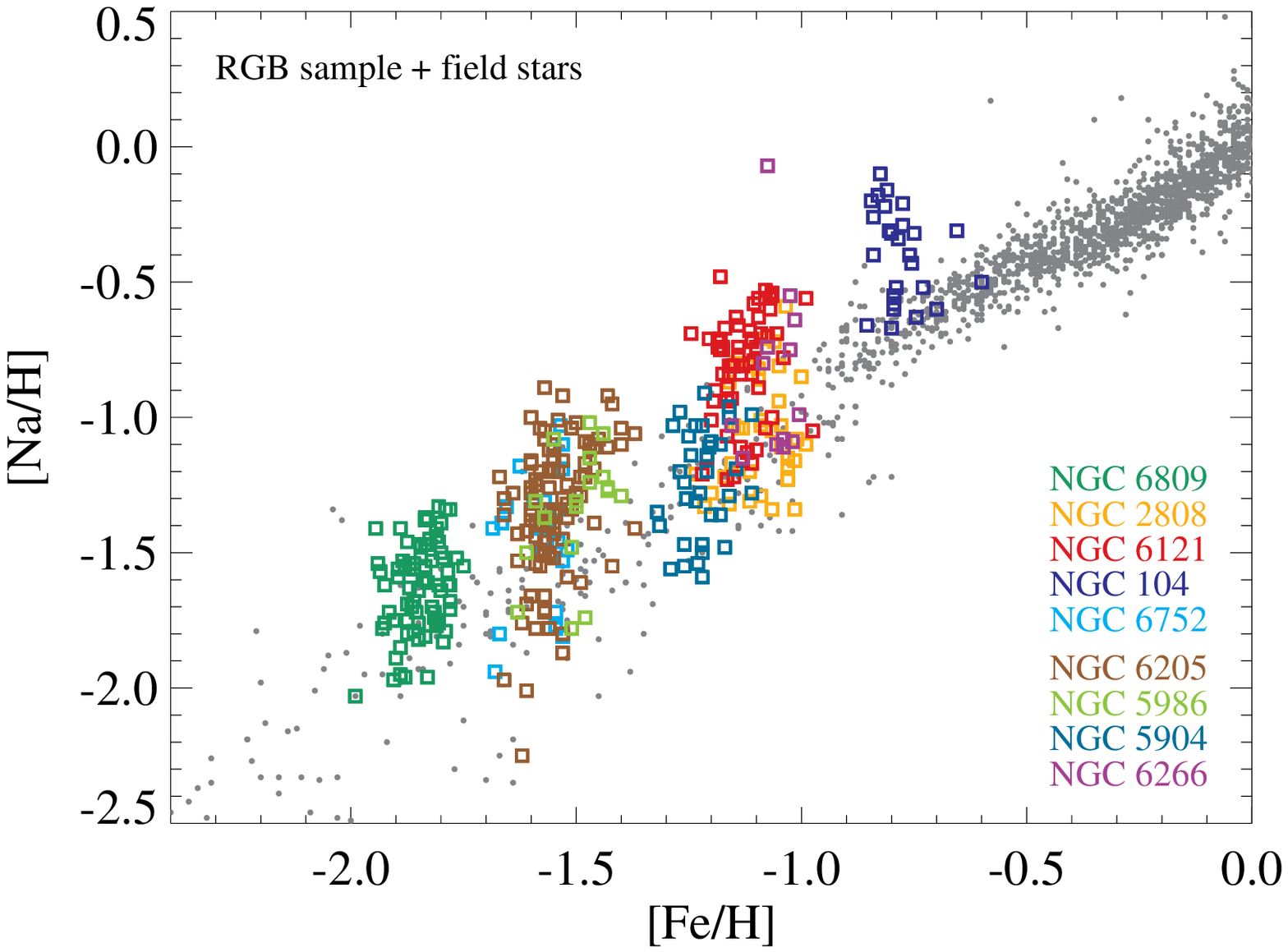}
    \includegraphics[width=0.47\textwidth]{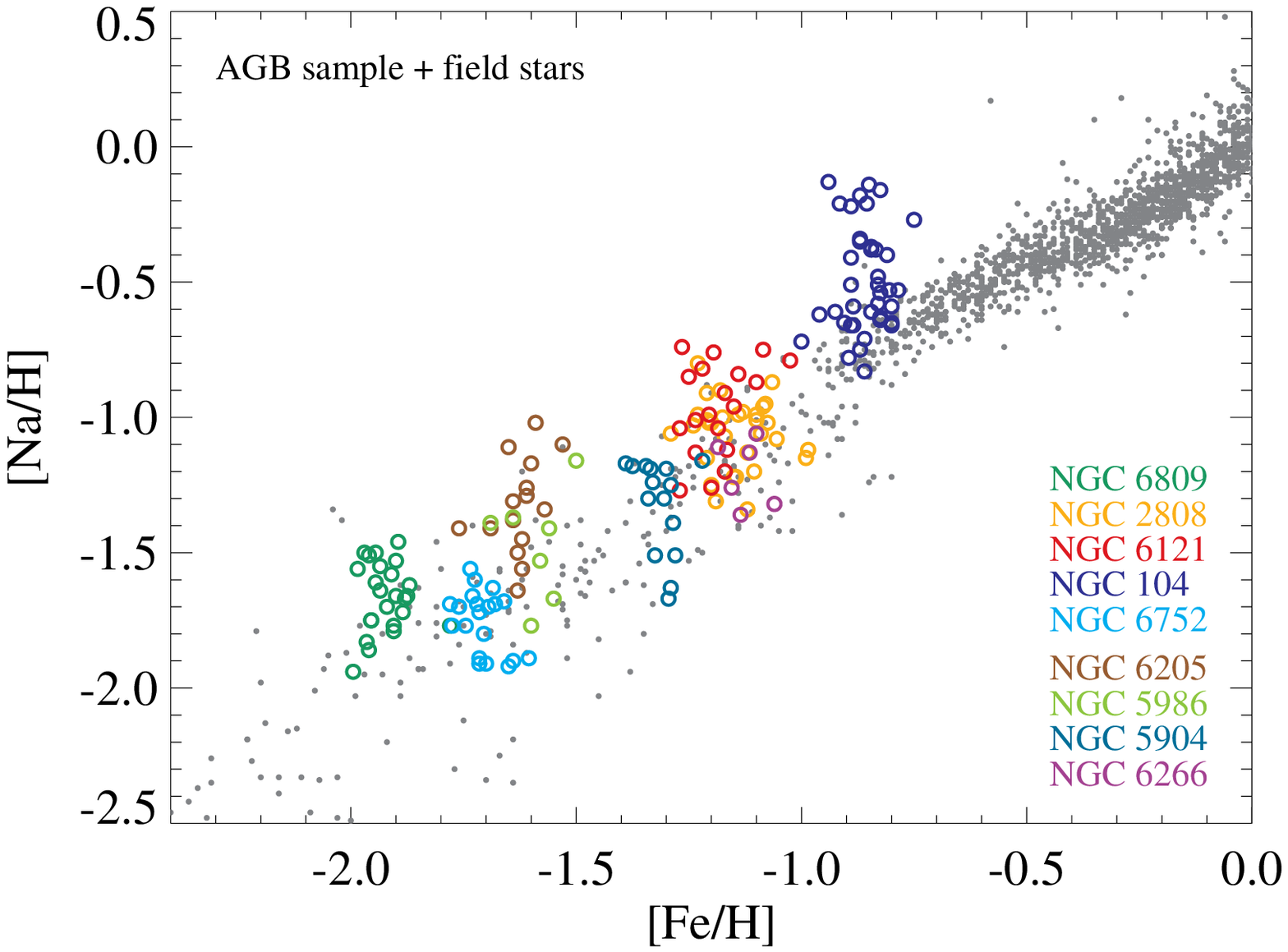}
     \caption{[Na/H] (NLTE) as a function of [Fe/H]
              for RGB (left) and AGB (right) stars in the nine GCs discussed in this paper (our analysis for NGC\,104, NGC\,2808, NGC\,6121, 
              NGC\,6809, and NGC\,6752; literature data for NGC\,5904, NGC\,5986, NGC\,6205, and NGC\,6266) and in Galactic field stars 
              from Carretta (2013, gray points).
              } 
     \label{GCvsfield}
  \end{figure*}

The number ratio between 1P and 2P stars has been extensively used to constrain the models that aim at explaining the chemical properties 
of the stellar populations in GCs (e.g., \citealp{PrantzosCharbonnel2006,Carretta2010,Decressin2010,SchaererCharbonnel2011,Charbonnel2014,Larsen2014,
BastianLardo2015,KhalajBaumgardt2015}). 
As a general concept, 1P and 2P refer respectively to GC stars that present chemical abundances similar or different to those of field stars of similar 
metallicity. This can be seen in  Fig.~\ref{GCvsfield} where we compare the Na data for RGB and AGB stars that we use in the discussion (including our 
determinations and data from the literature) with the Na data in Galactic field stars gathered and homogenized by \citet{Carretta2013}.
Using the same definition of [Na/H]$_\mathrm{cri}$ as in Paper\,I (see Sect. \ref{subsection:criteria}), we calculate the fraction of 2P stars 
(f$_\mathrm{2P}$) both on the RGB and AGB in each individual GC (Table~\ref{9GCs_comp}). Our results agree with the finding of \citet{Carretta2009a} that 
the 2P RGB component is present in all clusters, with a fraction between $\sim 50\%$ and $75\%$.  
For the AGB, we confirm previous abundance studies that revealed a complex picture, with some clusters being almost devoid of 2P AGB stars (NGC\,6752, 
NGC\,6266), some having similar 2P fractions on the AGB to on the RGB (NGC\,2808, NGC\,5904, and NGC\,6809), and all the possible intermediate cases 
between these two extreme behaviors. Overall, the fraction of 2P AGB stars varies between $\sim 0\%$ and $87\%$ (these extreme values corresponding to NGC\,6266 and NGC\,6205, respectively).

Obviously, the actual numbers one obtains for the fractions of 1P and 2P stars in a given GC depend on the adopted distinction criteria. 
For example, our conclusion that NGC\,6121 hosts 2P AGB stars seems at first sight to contradict the claim by \citet{MacLean2016} that 
there is no 2P AGB star in this cluster. 
However, \citet{MacLean2016} derived the population separation point (PSP) for their NGC\,6121 sample ([Na/O]\,=\,$-$0.16\,dex) by identifying a minimum 
in the [Na/O] distribution between the two subpopulations in the RGB sample of \citet{Marino2008}, and they found that the abundance distribution was 
consistent with all the AGB stars being of 1P, while 45\% of the RGB stars belong to 2P. 
However, if the criterion by \citet{Carretta2009a} to distinguish 1P and 2P stars (i.e., [Na/Fe]$_\mathrm{cri} =\,$[Na/Fe]$_\mathrm{min}\,+\,0.3$\,dex) 
is applied, one finds 6 out of 15 AGB stars and 65 out of 106 RGB stars belonging to 2P, which accounts for 40\% and 61\% of their AGB and RGB sample, 
respectively. Although it is still slightly different from the fractions found for our own sample (53\% AGB and 76\% RGB stars belonging to 2P), it is 
possible to identify some 2P AGB stars from the Na abundance distributions of both samples (ours and \citealp{MacLean2016}) when the same 1P$-$2P 
separation criterion is adopted. 
Moreover, from the [O/Fe]$-$[Na/Fe] distribution of the AGB stars studied by \citet{Marino2017} we can infer that the Na-rich/O-poor stars account for 
47\%, which is close to our 2P AGB fraction of 53\%. The photometric studies of \citet{Marino2017} and \citet{Lardo2017} also support the claim that NGC\,6121 
hosts multiple populations on AGB. Thus we believe that this cluster should have more than one stellar population along its AGB.

\subsection{Dependencies of the AGB population fraction and theoretical considerations}
\label{subsection:F2ptheory}

\subsubsection{Dependencies of the AGB population fraction on the GC parameters}
\label{subsubsection:dependency}

\begin{figure}[ht]
    \centering
     \includegraphics[width=0.45\textwidth]{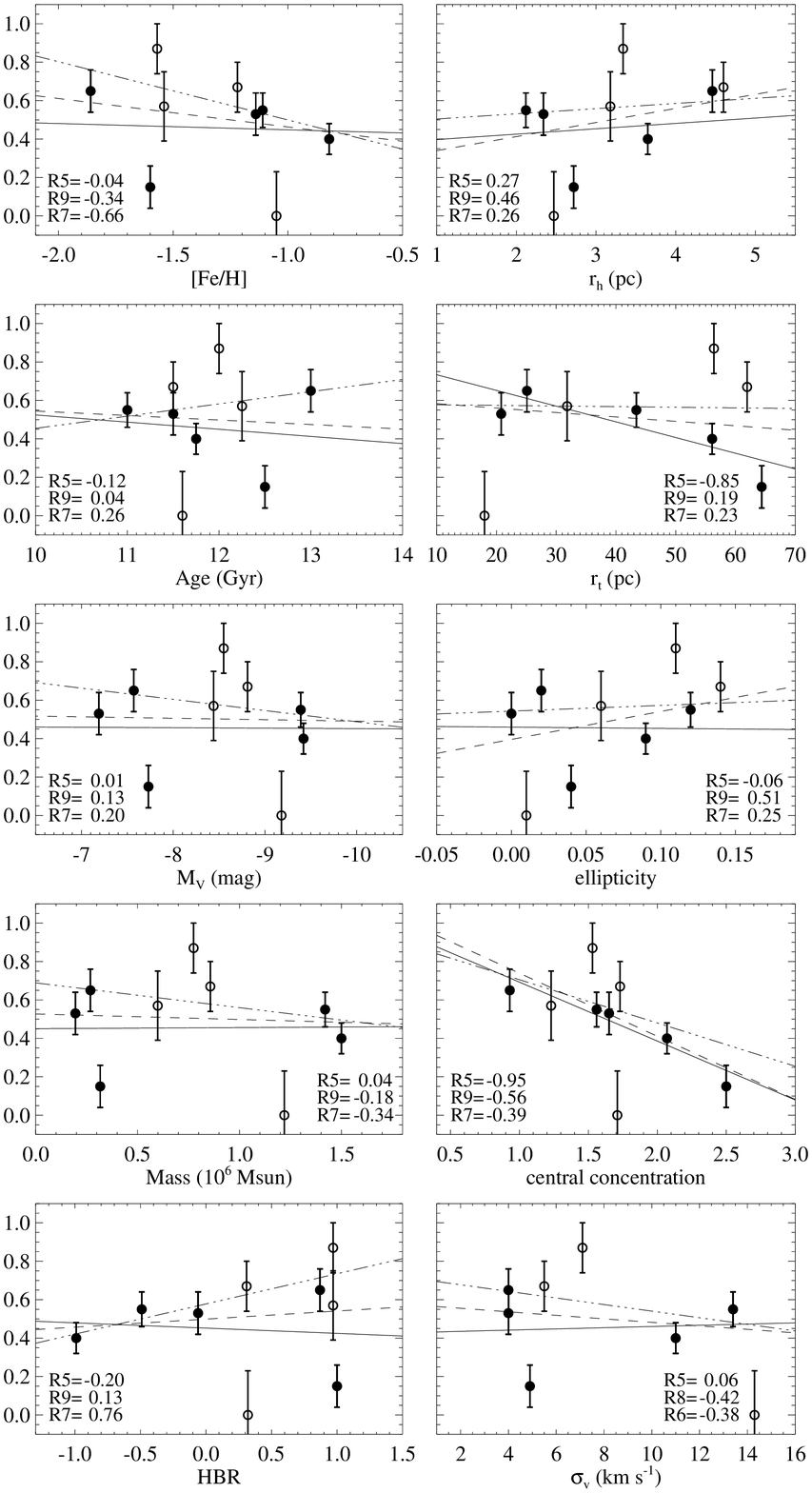}
     \caption{Relations between the fraction of 2P AGB stars (f$_\mathrm{2P,AGB}$) and the selected GC global parameters. 
              The data points were all derived using the criteria defined in Sect. \ref{subsection:criteria}. 
              Filled and open circles are used, respectively, for the five GCs we analyzed self-consistently (NGC\,6809, NGC\,2808, NGC\,6121, 
              NGC\,104, and NGC\,6752) and for those we took from the literature (NGC\,6205, NGC\,5986, NGC\,5904, and NGC\,6266).    
              The associated error bars are computed from Beta distribution. 
              The lines are the linear fits derived by least square fitting (with the errors taken into account); the solid and dashed lines 
              are fitted from our own five GCs and all the nine GCs, respectively, while the dash-dotted ones represent the fits disregarding 
              the two most scattered points of NGC\,6266 and NGC\,6752. 
              The Pearson correlation coefficients considering the five (R5), nine (R9), and seven (R7) data points,  respectively, are listed 
              at the bottom of each panel (except for the very bottom right panel, which reports the five (R5), eight (R8), and six (R6) data points, 
              as no $\sigma_\mathrm{v}$ data is available for NGC\,5986).
              }
     \label{relations_f2P_GCs}
  \end{figure}

We investigate possible dependencies between the fraction of 2P AGB stars (f$_\mathrm{2P,AGB}$) we derived and the GC global parameters listed 
in Table \ref{9GCs_comp}. 
We consider [Fe/H], age, absolute V magnitude (M$_\mathrm{V}$), mass, HB morphology index (HBR), 
half-light radius (r$_\mathrm{h}$), tidal radius (r$_\mathrm{t}$), ellipticity, central concentration (c.c.), and central velocity dispersion ($\sigma_\mathrm{v}$).
Figure \ref{relations_f2P_GCs} shows the data points and the linear fits derived by least square fitting with the errors taken into account. 
The Pearson correlation coefficients for each set of data points are also listed as a reference. 
When we consider only the five GCs that we have analyzed in a consistent manner (black points and solid lines), 
we obtain weak positive correlations between f$_\mathrm{2P,AGB}$ and half-light radius; 
and anticorrelations with GC age (weak), tidal radius, and central concentration. 
When taking all nine clusters together into account (dashed lines), 
weak positive relations exist between f$_\mathrm{2P,AGB}$ and half-light radius and ellipticity; 
and negative relations are shown between f$_\mathrm{2P,AGB}$ and [Fe/H] (weak), central concentration, and central velocity dispersion (weak). 
If the two extreme GCs showing almost no 2P AGB stars (NGC\,6752 and NGC\,6266) are disregarded (dash-dotted lines), 
f$_\mathrm{2P,AGB}$ correlates positively with GC age, HBR, and half-light radius (weak); 
while the dependencies on [Fe/H], M$_\mathrm{V}$, mass, central concentration, and central velocity dispersion are negative. 
Overall, only the anticorrelation between f$_\mathrm{2P,AGB}$ and central concentration agrees among the fits of all three subsamples. 
Except for central concentration, the inconsistent results of the three sets of fits indicate that there is no robust dependency for 
f$_\mathrm{2P,AGB}$ on any other global GC properties.
  
\subsubsection{Theoretical considerations}
\label{subsubsection:theoretical}

 \begin{figure}
    \centering
     \includegraphics[width=0.45\textwidth]{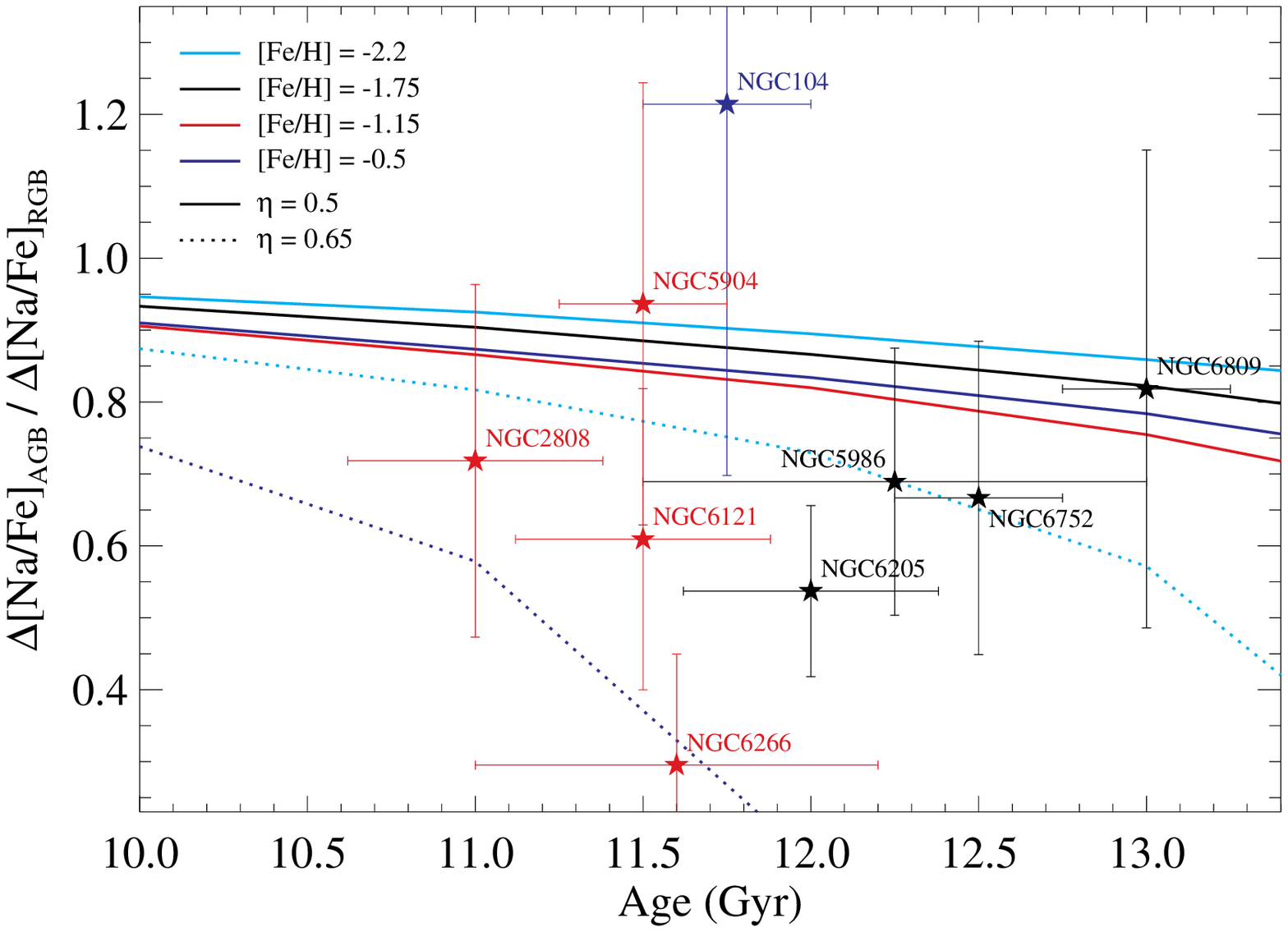}
       \caption{Ratio of the [Na/Fe] spreads on the AGB and RGB as a function of GC age. 
                Predictions of \citet{CCWC2016} for the original FRMS scenario are shown for different values of metallicity (colors) 
                and for two values of the $\eta$ parameter adopted for Reimers mass loss along the RGB (0.5 and 0.65, solid and dotted 
                lines respectively). The values derived in the present analysis are shown for the nine GCs (blue, red, and black stars, 
                with the colors of the symbols indicating the theoretical track that has the closest [Fe/H] from that of each individual GC).
                } 
     \label{NaFe_Obs-Model}
  \end{figure}

\citet{Charbonnel2013} proposed that the lack of sodium-rich AGB stars in NGC\,6752 could be due to a correlation between Na and He enrichment in 
the initial mixture of the low-mass stars we observe today, as predicted in the original FRMS scenario \citep{Decressin2007a}. They showed that in 
this framework, 2P stars born with an initial He abundance above a cutoff value (and consequently with an initial Na abundance above a cutoff value) 
are predicted to miss the AGB and evolve directly towards the white dwarf stage after central He burning, because of the impact of helium on stellar 
evolution. \citet{CCWC2016} showed that when one assumes the same initial Na$-$He correlation for all GCs within the FRMS framework as well as 
standard mass-loss rates on the RGB, the maximum Na content expected for 2P stars on the AGB is a function of both the metallicity and the age of GCs. 
At a given [Fe/H], younger clusters are expected to host AGB stars exhibiting a larger Na spread than older clusters; and, at a given age, higher Na 
dispersion along the AGB is predicted in metal-poor GCs than in the metal-rich ones. This is depicted in Fig.~\ref{NaFe_Obs-Model} where we show the 
model predictions for the ratio between the Na spread on the AGB and that on the RGB as a function of GC age and for different metallicities.

Among the nine currently available GCs, we have two pairs of GCs with similar metallicities, that is, NGC\,2808 and NGC\,6266, and NGC\,2808 and 
NGC\,6121, with the relatively younger GC in each pair always having larger [Na/H] dispersion and 2P-AGB fraction. 
NGC\,5986, NGC\,6205, and NGC\,6752 have comparable metallicities and slightly different ages, but still the relatively younger GC has more Na-rich 
2P AGB stars. Besides, NGC\,5904, NGC\,6121, and NGC\,6266 have almost the same age but slightly different metallicities, with the most metal-poor 
one showing the largest number of 2P AGB stars.
In other words, we have five GCs that lie in the domain where the AGB stars are expected to present large Na abundance dispersions within the FRMS 
framework. These are NGC\,104 and NGC\,2808 that are relatively young and metal-rich, NGC\,6205 and NGC\,5986 that have medium age and 
metallicity, and NGC\,6809 which is old and metal-poor. On the other hand, according to the FRMS prediction and based on their metallicities and ages, 
NGC\,5904, NGC\,6121, NGC\,6752, and NGC\,6266 are expected to have their AGB stars showing smaller Na abundance dispersions compared to their RGB 
counterparts. Thus, and to the very first order, the observations of these subgroups of GCs are consistent with the trends with age and metallicity 
predicted by the FRMS scenario. 

However, there are several important issues that might blur the general trends one tries to identify (see also the discussion in \citealt{CCWC2016}).
First, \citet{Charbonnel2013} and \citet{CCWC2016} showed that the maximum Na abundance expected on the AGB for a given age and metallicity strongly 
depends on the mass-loss rate adopted on the RGB (see also \citealp{Cassisi2014b}), especially for the oldest GCs and the most metal-rich ones. 
This is clear from Fig.~\ref{NaFe_Obs-Model} where we show the predictions for the ratio of the Na spreads on the AGB and RGB for two values (0.5 and 
0.65) of the $\eta$ parameter adopted in  Reimers prescription used by \citet{CCWC2016} to compute the mass-loss rate on the RGB. The trends with age 
and metallicity become stronger when higher mass-loss rates are considered. 
Additionally, the comparison between the predictions and the observed data for the nine GCs plotted in Fig.~\ref{NaFe_Obs-Model} supports variations 
of the RGB mass-loss efficiency from one GC to another within the range derived for $\eta$ by \citet[median and maximum values of 0.477 and 0.65 respectively]{McDonaldZijlstra2015} 
to explain the variations of the horizontal branch morphology for 56 Galactic GCs.
 
Second, both the observed and theoretical trends with age and metallicity are very weak (and we showed that in the observational case they depend on 
the considered subsamples), and scatters and exceptions exist in the data as can be seen in Fig.~\ref{relations_f2P_GCs}. This indicates to some extent 
that the fraction of 2P AGB stars is not simply affected by one single parameter/factor. Mass loss is certainly an important player as discussed above.

Third, although the ages we use here are taken from a single source (except for NGC\,6266), they were not derived for the [Fe/H] values reported in 
the present paper, which might also introduce some confusion in the derivation of the trends with age and metallicity. 

Finally, the initial Na$-$He correlation and its dependency with cluster properties is an important parameter to predict the theoretical value of the 
maximum Na expected on the AGB for a given GC. Unfortunately, this correlation is not constrained yet, which forced \citet{CCWC2016} to assume a similar 
relation over the metallicity range they investigated. 
To date, the only GCs for which non-negligible He enrichment were estimated using both photometry and spectroscopic measurements are NGC\,2808 \citep{Milone2015b} 
and NGC\,6266 \citep{Milone2015}. However, their Na distributions on the AGB are radically different and more similar to other GCs (NGC\,6121 and 
NGC\,6752, respectively) for which very modest He variations have been estimated (\citealp{Villanova2012} for NGC\,6121, \citealp{Milone2013} for NGC\,6752, 
and \citealp{Nardiello2015a} for both GCs). Therefore, the theoretical interpretation of the data is not straightforward, bearing also in mind that the 
above-mentioned comparisons depend on the chosen abundance data set (cf. abundance discrepencies between different studies mentioned earlier on). 

Overall, and as already pointed out in a more general context by \citet{Bastian2015}, the broad variety of chemical patterns certainly reveals a high 
degree of stochasticity that is challenging our understanding of the formation and evolution of GCs and of their stellar populations. 

\section{Summary}
\label{section:summary}
 
After the claim by C13 that no Na-rich AGB stars were present in NGC\,6752, several studies were recently devoted to the determination 
of the Na content of AGB stars in GCs. In this work, we present the first analysis for NGC\,6809, together with that of a large stellar sample in NGC\,104 
and NGC\,6121, and a re-analysis of NGC\,6752. This is complementary to our previous study of NGC\,2808 (Paper\,I). 
For these five GCs, we use a photometric method to derive effective temperature and surface gravity of the sample stars. Equivalent widths of unblended 
iron lines are measured to determine [Fe/H], and the Na abundances are derived via spectrum synthesis, assuming local thermodynamic equilibrium (LTE). 
Non-LTE corrections are then applied to both \ion{Fe}{i} and Na abundances, following the prescriptions by \citet{Lind2011}, \citet{Bergemann2012}, 
and \citet{Lind2012}. 
We provide a large set of stellar parameters and Na and Fe abundances derived self-consistently for the sample of 254 RGB stars and 145 AGB stars 
in five GCs. We compare our results with those of the literature, and we include in our discussion the only four other GCs (NGC\,5904, NGC\,5986, 
NGC\,6205, and NGC\,6266) for which Na has been studied along the AGB. The total sample of nine clusters covers a large range of [Fe/H] 
($-$1.86 to $-$0.82), age (11 to 13 Gyr), and global properties. 

We chose to use the [Na/H]$_\mathrm{NLTE}$ data to avoid the influence of the FeI-FeII discrepancy that has different behaviors in RGB and AGB stars, 
thus hampering the analysis. In addition, we define as Na-rich stars, or 2P stars, the objects that have a [Na/H] value above 
[Na/H]$_\mathrm{cri}$ =\,[Na/H]$_\mathrm{min}$\,+\,0.3\,dex, where [Na/H]$_\mathrm{min}$ is the minimum Na value derived for the sample (RGB + AGB) 
in a given cluster and 0.3\,dex is about one third of the [Na/H] spread. 
This definition agrees well with the general concept that 2P GC stars have higher Na abundance than field stars of similar metallicity. 
Although different criteria as well as abundance indicators (e.g., [Na/Fe] combined with O abundances, \citealt{Lapenna2016}) can obviously lead to 
different evaluations of the ratios between 1P and 2P stars, we chose this approach to carry out a consistent analysis of the Na abundance for our 
large sample of RGB and AGB stars in the nine GCs.

Our analysis confirms the complex picture of the Na abundances in the AGB stellar populations in Galactic GCs.  
Eight out of nine GCs under scrutiny exhibit a lower [Na/H] spread on the AGB than on the RGB, with the maximum [Na/H] abundance being consistently 
lower on the AGB than on the RGB. The only exception is the most metal-rich cluster NGC\,104, whose AGB stars show slightly larger Na spread than 
its RGB ones. Two clusters, NGC\,6752 and NGC\,6266, clearly stand out by showing (almost) no Na-rich (2P) AGB stars among the sample stars analyzed 
so far. We note, however, that uncertainties still exist since {\it i.} different conclusions have been reported for NGC\,6752 thus demanding further 
homogeneous studies in both AGB and RGB stars; and {\it ii.} no firm conclusion can be drawn for NGC\,6266, due to the paucity of stars scrutinized 
so far. We then find all possible behaviors between these two extremes. 
In NGC\,6121, NGC\,5904, and NGC\,2808, the AGB stars occupy about the bottom two thirds of the Na distribution of the RGB stars, with a difference 
of 0.26, 0.25, and 0.21~dex respectively between the maximum [Na/H] value of the AGB and RGB samples, indicating a deficit of very Na-rich AGB stars 
in these clusters. 
In the case of NGC\,5986, instead, the AGB and RGB samples show comparable spreads in [Na/H] with the maximum Na abundance of AGB stars being 
0.14\,dex lower than their RGB counterparts, but the AGB sample is still rather limited, thus weakening any drawn conclusion. 
In NGC\,6809 the Na abundance of the RGB sample spreads more evenly while the AGB stars tend to be more concentrated, and a 0.12\,dex lower 
[Na/H]$_\mathrm{max}$ present in the AGB sample compared to that of the RGB one. Finally, NGC\,6205 seems to have very few 1P AGB stars, and the 
maximum Na values for its RGB and AGB components are marginally consistent within the observational errors. 

Linear fits between the fraction of Na-rich 2P AGB stars and the GC parameters reveal that the AGB 2P fraction slightly anticorrelates with GC 
central concentration, with no conclusive results on possible trends with other GC parameters since they depend on the considered subsamples. 
By checking the AGB populations of pairs/subgroups of GCs and the trend of AGB Na abundance distributions of the nine GCs, we find that the 
current data roughly support the theoretical prediction of the original FRMS scenario according to which the initial Na and He abundances were correlated 
in the original mixture of the GC stars we observe today. 
However and as underlined in the discussion, this cannot be considered as a strong conclusion. Indeed, the predictions for the evolution of the 
stellar models along the AGB strongly depend on the mass loss on the RGB, and the initial Na$-$He correlation and its possible dependency with cluster 
properties is not sufficiently constrained yet. 
The fact that both the observed and the theoretical trends with age, metallicity, and other global GC properties are mostly very weak, and that scatter 
and exceptions exist indicate that the fraction of 2P AGB stars is affected by more than one or two factors and is probably subject to stochasticity.

\begin{acknowledgements}

YW acknowledges the support from the European Southern Observatory, via its ESO Studentship programme. 
This work was partly funded by the National Natural Science Foundation of China under grants 1233004 and 11390371, 
as well as the Strategic Priority Research Program The Emergence of Cosmological Structures of the Chinese Academy
of Sciences, Grant No. XDB09000000. 
CC and WC acknowledge support from the Swiss National Science Foundation (FNS) for the project 200020-159543 
Multiple stellar populations in massive star clusters - Formation, evolution, dynamics, impact on galactic evolution. 
We are indebted to Peter Stetson for kindly providing us with accurate Johnson-Morgan photometry. 
We thank Simon Campbell and Karin Lind for useful discussions, the latter also for giving us access to her NLTE correction grids. 
We thank the International Space Science Institute (ISSI, Bern, CH) for welcoming the activities of ISSI Team 271 
Massive star clusters across the Hubble Time (2013 - 2016). 
Finally, we thank the anonymous referees for their useful comments. 
This work has made use of the VALD database, operated at Uppsala University, the Institute of Astronomy RAS in Moscow, 
and the University of Vienna. 

\end{acknowledgements}



\begin{thebibliography}{108}
\expandafter\ifx\csname natexlab\endcsname\relax\def\natexlab#1{#1}\fi

\bibitem[{{Alonso} {et~al.}(1999){Alonso}, {Arribas}, \&
  {Mart{\'{\i}}nez-Roger}}]{Alonso1999}
{Alonso}, A., {Arribas}, S., \& {Mart{\'{\i}}nez-Roger}, C. 1999, \aaps, 140,
  261

\bibitem[{{Anderson} {et~al.}(2009){Anderson}, {Piotto}, {King}, {Bedin}, \&
  {Guhathakurta}}]{Anderson2009}
{Anderson}, J., {Piotto}, G., {King}, I.~R., {Bedin}, L.~R., \& {Guhathakurta},
  P. 2009, \apjl, 697, L58

\bibitem[{{Asplund} {et~al.}(2009){Asplund}, {Grevesse}, {Sauval}, \&
  {Scott}}]{Asplund2009}
{Asplund}, M., {Grevesse}, N., {Sauval}, A.~J., \& {Scott}, P. 2009, \araa, 47,
  481

\bibitem[{{Bastian} {et~al.}(2015){Bastian}, {Cabrera-Ziri}, \&
  {Salaris}}]{Bastian2015}
{Bastian}, N., {Cabrera-Ziri}, I., \& {Salaris}, M. 2015, \mnras, 449, 3333

\bibitem[{{Bastian} {et~al.}(2013){Bastian}, {Lamers}, {de Mink}, {Longmore},
  {Goodwin}, \& {Gieles}}]{Bastian2013}
{Bastian}, N., {Lamers}, H.~J.~G.~L.~M., {de Mink}, S.~E., {et~al.} 2013,
  \mnras, 436, 2398

\bibitem[{{Bastian} \& {Lardo}(2015)}]{BastianLardo2015}
{Bastian}, N. \& {Lardo}, C. 2015, \mnras, 453, 357

\bibitem[{{Bergemann} {et~al.}(2012){Bergemann}, {Lind}, {Collet}, {Magic}, \&
  {Asplund}}]{Bergemann2012}
{Bergemann}, M., {Lind}, K., {Collet}, R., {Magic}, Z., \& {Asplund}, M. 2012,
  \mnras, 427, 27

\bibitem[{{Bonatto} {et~al.}(2013){Bonatto}, {Campos}, \&
  {Kepler}}]{Bonatto2013}
{Bonatto}, C., {Campos}, F., \& {Kepler}, S.~O. 2013, \mnras, 435, 263

\bibitem[{{Bono} {et~al.}(2008){Bono}, {Stetson}, {Sanna}, {Piersimoni},
  {Freyhammer}, {Bouzid}, {Buonanno}, {Calamida}, {Caputo}, {Corsi}, {Di
  Cecco}, {Dall'Ora}, {Ferraro}, {Iannicola}, {Monelli}, {Nonino}, {Pulone},
  {Sterken}, {Storm}, {Tuvikene}, \& {Walker}}]{Bono2008}
{Bono}, G., {Stetson}, P.~B., {Sanna}, N., {et~al.} 2008, \apjl, 686, L87

\bibitem[{{Boyles} {et~al.}(2011){Boyles}, {Lorimer}, {Turk}, {Mnatsakanov},
  {Lynch}, {Ransom}, {Freire}, \& {Belczynski}}]{Boyles2011}
{Boyles}, J., {Lorimer}, D.~R., {Turk}, P.~J., {et~al.} 2011, \apj, 742, 51

\bibitem[{{Campbell} {et~al.}(2013){Campbell}, {D'Orazi}, {Yong},
  {Constantino}, {Lattanzio}, {Stancliffe}, {Angelou}, {Wylie-de Boer}, \&
  {Grundahl}}]{Campbell2013}
{Campbell}, S.~W., {D'Orazi}, V., {Yong}, D., {et~al.} 2013, \nat, 498, 198
 
\bibitem[{{Campbell} {et~al.}(2017){Campbell}, {MacLean}, {D'Orazi},
  {Casagrande}, {de Silva}, {Yong}, {Cottrell}, \& {Lattanzio}}]{Campbell2017}
{Campbell}, S.~W., {MacLean}, B.~T., {D'Orazi}, V., {et~al.} 2017, ArXiv
  e-prints [\eprint[arXiv]{1707.02840}]

\bibitem[{{Cardelli} {et~al.}(1989){Cardelli}, {Clayton}, \&
  {Mathis}}]{Cardelli1989}
{Cardelli}, J.~A., {Clayton}, G.~C., \& {Mathis}, J.~S. 1989, \apj, 345, 245

\bibitem[{{Carretta}(2013)}]{Carretta2013}
{Carretta}, E. 2013, \aap, 557, A128

\bibitem[{{Carretta}(2014)}]{Carretta2014b}
{Carretta}, E. 2014, \apjl, 795, L28

\bibitem[{{Carretta}(2016)}]{Carretta2016}
{Carretta}, E. 2016, ArXiv e-prints [\eprint[arXiv]{1611.04728}]

\bibitem[{{Carretta} {et~al.}(2014){Carretta}, {Bragaglia}, {Gratton},
  {D'Orazi}, {Lucatello}, {Momany}, {Sollima}, {Bellazzini}, {Catanzaro}, \&
  {Leone}}]{Carretta2014a}
{Carretta}, E., {Bragaglia}, A., {Gratton}, R.~G., {et~al.} 2014, \aap, 564,
  A60

\bibitem[{{Carretta} {et~al.}(2006){Carretta}, {Bragaglia}, {Gratton}, {Leone},
  {Recio-Blanco}, \& {Lucatello}}]{Carretta2006}
{Carretta}, E., {Bragaglia}, A., {Gratton}, R.~G., {et~al.} 2006, \aap, 450,
  523

\bibitem[{{Carretta} {et~al.}(2009){Carretta}, {Bragaglia}, {Gratton},
  {Lucatello}, {Catanzaro}, {Leone}, {Bellazzini}, {Claudi}, {D'Orazi},
  {Momany}, {Ortolani}, {Pancino}, {Piotto}, {Recio-Blanco}, \&
  {Sabbi}}]{Carretta2009a}
{Carretta}, E., {Bragaglia}, A., {Gratton}, R.~G., {et~al.} 2009, \aap, 505,
  117

\bibitem[{{Carretta} {et~al.}(2010){Carretta}, {Bragaglia}, {Gratton},
  {Recio-Blanco}, {Lucatello}, {D'Orazi}, \& {Cassisi}}]{Carretta2010}
{Carretta}, E., {Bragaglia}, A., {Gratton}, R.~G., {et~al.} 2010, \aap, 516,
  A55

\bibitem[{{Carretta} {et~al.}(2005){Carretta}, {Gratton}, {Lucatello},
  {Bragaglia}, \& {Bonifacio}}]{Carretta2005}
{Carretta}, E., {Gratton}, R.~G., {Lucatello}, S., {Bragaglia}, A., \&
  {Bonifacio}, P. 2005, \aap, 433, 597

\bibitem[{{Cassisi} \& {Salaris}(2014)}]{Cassisi2014a}
{Cassisi}, S. \& {Salaris}, M. 2014, \aap, 563, A10

\bibitem[{{Cassisi} {et~al.}(2014){Cassisi}, {Salaris}, {Pietrinferni}, {Vink},
  \& {Monelli}}]{Cassisi2014b}
{Cassisi}, S., {Salaris}, M., {Pietrinferni}, A., {Vink}, J.~S., \& {Monelli},
  M. 2014, \aap, 571, A81

\bibitem[{{Chantereau} {et~al.}(2015){Chantereau}, {Charbonnel}, \&
  {Decressin}}]{Chantereau2015}
{Chantereau}, W., {Charbonnel}, C., \& {Decressin}, T. 2015, \aap, 578, A117

\bibitem[{{Chantereau} {et~al.}(2016){Chantereau}, {Charbonnel}, \&
  {Meynet}}]{Chantereau2016}
{Chantereau}, W., {Charbonnel}, C., \& {Meynet}, G. 2016, \aap, 592, A111

\bibitem[{{Charbonnel}(2016)}]{Charbonnel2016EAS}
{Charbonnel}, C. 2016, in EAS Publications Series, Vol.~80, EAS Publications
  Series, ed. E.~{Moraux}, Y.~{Lebreton}, \& C.~{Charbonnel}, 177--226

\bibitem[{{Charbonnel} \& {Chantereau}(2016)}]{CCWC2016}
{Charbonnel}, C. \& {Chantereau}, W. 2016, \aap, 586, A21

\bibitem[{{Charbonnel} {et~al.}(2013){Charbonnel}, {Chantereau}, {Decressin},
  {Meynet}, \& {Schaerer}}]{Charbonnel2013}
{Charbonnel}, C., {Chantereau}, W., {Decressin}, T., {Meynet}, G., \&
  {Schaerer}, D. 2013, \aap, 557, L17

\bibitem[{{Charbonnel} {et~al.}(2014){Charbonnel}, {Chantereau}, {Krause},
  {Primas}, \& {Wang}}]{Charbonnel2014}
{Charbonnel}, C., {Chantereau}, W., {Krause}, M., {Primas}, F., \& {Wang}, Y.
  2014, \aap, 569, L6

\bibitem[{{Cordero} {et~al.}(2014){Cordero}, {Pilachowski}, {Johnson},
  {McDonald}, {Zijlstra}, \& {Simmerer}}]{Cordero2014}
{Cordero}, M.~J., {Pilachowski}, C.~A., {Johnson}, C.~I., {et~al.} 2014, \apj,
  780, 94

\bibitem[{{Cudworth} \& {Rees}(1990)}]{Cudworth1990}
{Cudworth}, K.~M. \& {Rees}, R. 1990, \aj, 99, 1491

\bibitem[{{D'Antona} {et~al.}(2010){D'Antona}, {Ventura}, {Caloi}, {D'Ercole},
  {Vesperini}, {Carini}, \& {Di Criscienzo}}]{DAntona2010}
{D'Antona}, F., {Ventura}, P., {Caloi}, V., {et~al.} 2010, \apjl, 715, L63

\bibitem[{{de Mink} {et~al.}(2009){de Mink}, {Pols}, {Langer}, \&
  {Izzard}}]{deMink2009}
{de Mink}, S.~E., {Pols}, O.~R., {Langer}, N., \& {Izzard}, R.~G. 2009, \aap,
  507, L1

\bibitem[{{Decressin} {et~al.}(2010){Decressin}, {Baumgardt}, {Charbonnel}, \&
  {Kroupa}}]{Decressin2010}
{Decressin}, T., {Baumgardt}, H., {Charbonnel}, C., \& {Kroupa}, P. 2010, \aap,
  516, A73

\bibitem[{{Decressin} {et~al.}(2007{\natexlab{a}}){Decressin}, {Charbonnel}, \&
  {Meynet}}]{Decressin2007b}
{Decressin}, T., {Charbonnel}, C., \& {Meynet}, G. 2007{\natexlab{a}}, \aap,
  475, 859

\bibitem[{{Decressin} {et~al.}(2007{\natexlab{b}}){Decressin}, {Meynet},
  {Charbonnel}, {Prantzos}, \& {Ekstr{\"o}m}}]{Decressin2007a}
{Decressin}, T., {Meynet}, G., {Charbonnel}, C., {Prantzos}, N., \&
  {Ekstr{\"o}m}, S. 2007{\natexlab{b}}, \aap, 464, 1029

\bibitem[{{Denissenkov} \& {Hartwick}(2014)}]{Denissenkov2014}
{Denissenkov}, P.~A. \& {Hartwick}, F.~D.~A. 2014, \mnras, 437, L21

\bibitem[{{Denissenkov} {et~al.}(2015){Denissenkov}, {VandenBerg}, {Hartwick},
  {Herwig}, {Weiss}, \& {Paxton}}]{Denissenkov2015}
{Denissenkov}, P.~A., {VandenBerg}, D.~A., {Hartwick}, F.~D.~A., {et~al.} 2015,
  \mnras, 448, 3314

\bibitem[{{D'Ercole} {et~al.}(2010){D'Ercole}, {D'Antona}, {Ventura},
  {Vesperini}, \& {McMillan}}]{DErcole2010}
{D'Ercole}, A., {D'Antona}, F., {Ventura}, P., {Vesperini}, E., \& {McMillan},
  S.~L.~W. 2010, \mnras, 407, 854

\bibitem[{{di Criscienzo} {et~al.}(2010){di Criscienzo}, {Ventura}, {D'Antona},
  {Milone}, \& {Piotto}}]{diCriscienzo2010}
{di Criscienzo}, M., {Ventura}, P., {D'Antona}, F., {Milone}, A., \& {Piotto},
  G. 2010, \mnras, 408, 999

\bibitem[{{Garc{\'{\i}}a-Hern{\'a}ndez}
  {et~al.}(2015){Garc{\'{\i}}a-Hern{\'a}ndez}, {M{\'e}sz{\'a}ros}, {Monelli},
  {Cassisi}, {Stetson}, {Zamora}, {Shetrone}, \&
  {Lucatello}}]{GarciaHernandez2015}
{Garc{\'{\i}}a-Hern{\'a}ndez}, D.~A., {M{\'e}sz{\'a}ros}, S., {Monelli}, M.,
  {et~al.} 2015, \apjl, 815, L4

\bibitem[{{Gratton} {et~al.}(2003){Gratton}, {Bragaglia}, {Carretta},
  {Clementini}, {Desidera}, {Grundahl}, \& {Lucatello}}]{Gratton2003}
{Gratton}, R.~G., {Bragaglia}, A., {Carretta}, E., {et~al.} 2003, \aap, 408,
  529

\bibitem[{{Gratton} {et~al.}(1999){Gratton}, {Carretta}, {Eriksson}, \&
  {Gustafsson}}]{Gratton1999}
{Gratton}, R.~G., {Carretta}, E., {Eriksson}, K., \& {Gustafsson}, B. 1999,
  \aap, 350, 955

\bibitem[{{Gruyters} {et~al.}(2017){Gruyters}, {Casagrande}, {Milone},
  {Hodgkin}, {Serenelli}, \& {Feltzing}}]{Gruyters2017}
{Gruyters}, P., {Casagrande}, L., {Milone}, A.~P., {et~al.} 2017, \aap, 603,
  A37
  
\bibitem[{{Gustafsson} {et~al.}(2008){Gustafsson}, {Edvardsson}, {Eriksson},
  {J{\o}rgensen}, {Nordlund}, \& {Plez}}]{Gustafsson2008}
{Gustafsson}, B., {Edvardsson}, B., {Eriksson}, K., {et~al.} 2008, \aap, 486,
  951

\bibitem[{{Harris}(1996)}]{Harris1996}
{Harris}, W.~E. 1996, \aj, 112, 1487

\bibitem[{{Hendricks} {et~al.}(2012){Hendricks}, {Stetson}, {VandenBerg}, \&
  {Dall'Ora}}]{Hendricks2012}
{Hendricks}, B., {Stetson}, P.~B., {VandenBerg}, D.~A., \& {Dall'Ora}, M. 2012,
  \aj, 144, 25

\bibitem[{{Ivans} {et~al.}(2001){Ivans}, {Kraft}, {Sneden}, {Smith}, {Rich}, \&
  {Shetrone}}]{Ivans2001}
{Ivans}, I.~I., {Kraft}, R.~P., {Sneden}, C., {et~al.} 2001, \aj, 122, 1438

\bibitem[{{Izzard} {et~al.}(2013){Izzard}, {de Mink}, {Pols}, {Langer}, {Sana},
  \& {de Koter}}]{Izzard2013}
{Izzard}, R.~G., {de Mink}, S.~E., {Pols}, O.~R., {et~al.} 2013, \memsai, 84,
  171

\bibitem[{{Johnson} {et~al.}(2017){Johnson}, {Caldwell}, {Rich}, {Mateo},
  {Bailey}, {Olszewski}, \& {Walker}}]{Johnson2017}
{Johnson}, C.~I., {Caldwell}, N., {Rich}, R.~M., {et~al.} 2017, \apj, 842, 24
  
\bibitem[{{Johnson} {et~al.}(2015){Johnson}, {McDonald}, {Pilachowski},
  {Mateo}, {Bailey}, {Cordero}, {Zijlstra}, {Crane}, {Olszewski}, {Shectman},
  \& {Thompson}}]{Johnson2015}
{Johnson}, C.~I., {McDonald}, I., {Pilachowski}, C.~A., {et~al.} 2015, \aj,
  149, 71

\bibitem[{{Johnson} \& {Pilachowski}(2012)}]{Johnson2012}
{Johnson}, C.~I. \& {Pilachowski}, C.~A. 2012, \apjl, 754, L38

\bibitem[{{Khalaj} \& {Baumgardt}(2015)}]{KhalajBaumgardt2015}
{Khalaj}, P. \& {Baumgardt}, H. 2015, \mnras, 452, 924

\bibitem[{{Krause} {et~al.}(2013){Krause}, {Charbonnel}, {Decressin}, {Meynet},
  \& {Prantzos}}]{Krause2013}
{Krause}, M., {Charbonnel}, C., {Decressin}, T., {Meynet}, G., \& {Prantzos},
  N. 2013, \aap, 552, A121

\bibitem[{{Krause} {et~al.}(2016){Krause}, {Charbonnel}, {Bastian}, \&
  {Diehl}}]{Krause2016}
{Krause}, M.~G.~H., {Charbonnel}, C., {Bastian}, N., \& {Diehl}, R. 2016, \aap,
  587, A53

\bibitem[{{Lai} {et~al.}(2011){Lai}, {Smith}, {Bolte}, {Johnson}, {Lucatello},
  {Kraft}, \& {Sneden}}]{Lai2011}
{Lai}, D.~K., {Smith}, G.~H., {Bolte}, M., {et~al.} 2011, \aj, 141, 62

\bibitem[{{Lapenna} {et~al.}(2016){Lapenna}, {Lardo}, {Mucciarelli}, {Salaris},
  {Ferraro}, {Lanzoni}, {Massari}, {Stetson}, {Cassisi}, \&
  {Savino}}]{Lapenna2016}
{Lapenna}, E., {Lardo}, C., {Mucciarelli}, A., {et~al.} 2016, \apjl, 826, L1

\bibitem[{{Lapenna} {et~al.}(2015){Lapenna}, {Mucciarelli}, {Ferraro},
  {Origlia}, {Lanzoni}, {Massari}, \& {Dalessandro}}]{Lapenna2015}
{Lapenna}, E., {Mucciarelli}, A., {Ferraro}, F.~R., {et~al.} 2015, \apj, 813,
  97

\bibitem[{{Lapenna} {et~al.}(2014){Lapenna}, {Mucciarelli}, {Lanzoni},
  {Ferraro}, {Dalessandro}, {Origlia}, \& {Massari}}]{Lapenna2014}
{Lapenna}, E., {Mucciarelli}, A., {Lanzoni}, B., {et~al.} 2014, \apj, 797, 124

\bibitem[{{Lardo} {et~al.}(2017){Lardo}, {Salaris}, {Savino}, {Donati},
  {Stetson}, \& {Cassisi}}]{Lardo2017}
{Lardo}, C., {Salaris}, M., {Savino}, A., {et~al.} 2017, \mnras, 466, 3507

\bibitem[{{Larsen} {et~al.}(2015){Larsen}, {Baumgardt}, {Bastian}, {Brodie},
  {Grundahl}, \& {Strader}}]{Larsen2015}
{Larsen}, S.~S., {Baumgardt}, H., {Bastian}, N., {et~al.} 2015, \apj, 804, 71

\bibitem[{{Larsen} {et~al.}(2014){Larsen}, {Brodie}, {Forbes}, \&
  {Strader}}]{Larsen2014}
{Larsen}, S.~S., {Brodie}, J.~P., {Forbes}, D.~A., \& {Strader}, J. 2014, \aap,
  565, A98

\bibitem[{{Lind} {et~al.}(2011){Lind}, {Asplund}, {Barklem}, \&
  {Belyaev}}]{Lind2011}
{Lind}, K., {Asplund}, M., {Barklem}, P.~S., \& {Belyaev}, A.~K. 2011, \aap,
  528, A103

\bibitem[{{Lind} {et~al.}(2012){Lind}, {Bergemann}, \& {Asplund}}]{Lind2012}
{Lind}, K., {Bergemann}, M., \& {Asplund}, M. 2012, \mnras, 427, 50

\bibitem[{{Mackey} \& {van den Bergh}(2005)}]{Mackey2005}
{Mackey}, A.~D. \& {van den Bergh}, S. 2005, \mnras, 360, 631

\bibitem[{{MacLean} {et~al.}(2016){MacLean}, {Campbell}, {De Silva},
  {Lattanzio}, {D'Orazi}, {Simpson}, \& {Momany}}]{MacLean2016}
{MacLean}, B.~T., {Campbell}, S.~W., {De Silva}, G.~M., {et~al.} 2016, \mnras,
  460, L69

\bibitem[{{Maeder} \& {Meynet}(2006)}]{MaederMeynet2006}
{Maeder}, A. \& {Meynet}, G. 2006, \aap, 448, L37

\bibitem[{{Marconi} {et~al.}(2015){Marconi}, {Coppola}, {Bono}, {Braga},
  {Pietrinferni}, {Buonanno}, {Castellani}, {Musella}, {Ripepi}, \&
  {Stellingwerf}}]{Marconi2015}
{Marconi}, M., {Coppola}, G., {Bono}, G., {et~al.} 2015, \apj, 808, 50

\bibitem[{{Marino} {et~al.}(2014){Marino}, {Milone}, {Przybilla}, {Bergemann},
  {Lind}, {Asplund}, {Cassisi}, {Catelan}, {Casagrande}, {Valcarce}, {Bedin},
  {Cort{\'e}s}, {D'Antona}, {Jerjen}, {Piotto}, {Schlesinger}, {Zoccali}, \&
  {Angeloni}}]{Marino2014}
{Marino}, A.~F., {Milone}, A.~P., {Przybilla}, N., {et~al.} 2014, \mnras, 437,
  1609

\bibitem[{{Marino} {et~al.}(2017){Marino}, {Milone}, {Yong}, {Da Costa},
  {Asplund}, {Bedin}, {Jerjen}, {Nardiello}, {Piotto}, {Renzini}, \&
  {Shetrone}}]{Marino2017}
{Marino}, A.~F., {Milone}, A.~P., {Yong}, D., {et~al.} 2017, \apj, 843, 66
  
\bibitem[{{Marino} {et~al.}(2008){Marino}, {Villanova}, {Piotto}, {Milone},
  {Momany}, {Bedin}, \& {Medling}}]{Marino2008}
{Marino}, A.~F., {Villanova}, S., {Piotto}, G., {et~al.} 2008, \aap, 490, 625

\bibitem[{{McDonald} \& {Zijlstra}(2015)}]{McDonaldZijlstra2015}
{McDonald}, I. \& {Zijlstra}, A.~A. 2015, \mnras, 448, 502

\bibitem[{{Milone}(2015)}]{Milone2015}
{Milone}, A.~P. 2015, \mnras, 446, 1672

\bibitem[{{Milone} {et~al.}(2013){Milone}, {Marino}, {Piotto}, {Bedin},
  {Anderson}, {Aparicio}, {Bellini}, {Cassisi}, {D'Antona}, {Grundahl},
  {Monelli}, \& {Yong}}]{Milone2013}
{Milone}, A.~P., {Marino}, A.~F., {Piotto}, G., {et~al.} 2013, \apj, 767, 120

\bibitem[{{Milone} {et~al.}(2015{\natexlab{a}}){Milone}, {Marino}, {Piotto},
  {Bedin}, {Anderson}, {Renzini}, {King}, {Bellini}, {Brown}, {Cassisi},
  {D'Antona}, {Jerjen}, {Nardiello}, {Salaris}, {Marel}, {Vesperini}, {Yong},
  {Aparicio}, {Sarajedini}, \& {Zoccali}}]{Milone2015a}
{Milone}, A.~P., {Marino}, A.~F., {Piotto}, G., {et~al.} 2015{\natexlab{a}},
  \mnras, 447, 927

\bibitem[{{Milone} {et~al.}(2015{\natexlab{b}}){Milone}, {Marino}, {Piotto},
  {Renzini}, {Bedin}, {Anderson}, {Cassisi}, {D'Antona}, {Bellini}, {Jerjen},
  {Pietrinferni}, \& {Ventura}}]{Milone2015b}
{Milone}, A.~P., {Marino}, A.~F., {Piotto}, G., {et~al.} 2015{\natexlab{b}},
  \apj, 808, 51

\bibitem[{{Milone} {et~al.}(2012{\natexlab{a}}){Milone}, {Piotto}, {Bedin},
  {Cassisi}, {Anderson}, {Marino}, {Pietrinferni}, \& {Aparicio}}]{Milone2012}
{Milone}, A.~P., {Piotto}, G., {Bedin}, L.~R., {et~al.} 2012{\natexlab{a}},
  \aap, 537, A77

\bibitem[{{Milone} {et~al.}(2012{\natexlab{b}}){Milone}, {Piotto}, {Bedin},
  {King}, {Anderson}, {Marino}, {Bellini}, {Gratton}, {Renzini}, {Stetson},
  {Cassisi}, {Aparicio}, {Bragaglia}, {Carretta}, {D'Antona}, {Di Criscienzo},
  {Lucatello}, {Monelli}, \& {Pietrinferni}}]{Milone2012b}
{Milone}, A.~P., {Piotto}, G., {Bedin}, L.~R., {et~al.} 2012{\natexlab{b}},
  \apj, 744, 58

\bibitem[{{Monelli} {et~al.}(2013){Monelli}, {Milone}, {Stetson}, {Marino},
  {Cassisi}, {del Pino Molina}, {Salaris}, {Aparicio}, {Asplund}, {Grundahl},
  {Piotto}, {Weiss}, {Carrera}, {Cebri{\'a}n}, {Murabito}, {Pietrinferni}, \&
  {Sbordone}}]{Monelli2013}
{Monelli}, M., {Milone}, A.~P., {Stetson}, P.~B., {et~al.} 2013, \mnras, 431,
  2126
  
\bibitem[{{Mucciarelli} {et~al.}(2015){Mucciarelli}, {Lapenna}, {Massari},
  {Ferraro}, \& {Lanzoni}}]{Mucciarelli2015}
{Mucciarelli}, A., {Lapenna}, E., {Massari}, D., {Ferraro}, F.~R., \&
  {Lanzoni}, B. 2015, \apj, 801, 69
  
\bibitem[{{Mucciarelli} {et~al.}(2011){Mucciarelli}, {Salaris}, {Lovisi},
  {Ferraro}, {Lanzoni}, {Lucatello}, \& {Gratton}}]{Mucciarelli2011}
{Mucciarelli}, A., {Salaris}, M., {Lovisi}, L., {et~al.} 2011, \mnras, 412, 81

\bibitem[{{Nardiello} {et~al.}(2015{\natexlab{a}}){Nardiello}, {Milone},
  {Piotto}, {Marino}, {Bellini}, \& {Cassisi}}]{Nardiello2015a}
{Nardiello}, D., {Milone}, A.~P., {Piotto}, G., {et~al.} 2015{\natexlab{a}},
  \aap, 573, A70

\bibitem[{{Nardiello} {et~al.}(2015{\natexlab{b}}){Nardiello}, {Piotto},
  {Milone}, {Marino}, {Bedin}, {Anderson}, {Aparicio}, {Bellini}, {Cassisi},
  {D'Antona}, {Hidalgo}, {Ortolani}, {Pietrinferni}, {Renzini}, {Salaris},
  {Marel}, \& {Vesperini}}]{Nardiello2015b}
{Nardiello}, D., {Piotto}, G., {Milone}, A.~P., {et~al.} 2015{\natexlab{b}},
  \mnras, 451, 312

\bibitem[{{Olech} {et~al.}(1999){Olech}, {Kaluzny}, {Thompson}, {Pych},
  {Krzeminski}, \& {Schwarzenberg-Czerny}}]{Olech1999}
{Olech}, A., {Kaluzny}, J., {Thompson}, I.~B., {et~al.} 1999, \aj, 118, 442

\bibitem[{{Pancino} {et~al.}(2010){Pancino}, {Rejkuba}, {Zoccali}, \&
  {Carrera}}]{Pancino2010}
{Pancino}, E., {Rejkuba}, M., {Zoccali}, M., \& {Carrera}, R. 2010, \aap, 524,
  A44

\bibitem[{{Pasquini} {et~al.}(2003){Pasquini}, {Alonso}, {Avila}, {Barriga},
  {Biereichel}, {Buzzoni}, {Cavadore}, {Cumani}, {Dekker}, {Delabre}, {Kaufer},
  {Kotzlowski}, {Hill}, {Lizon}, {Nees}, {Santin}, {Schmutzer}, {Kesteren}, \&
  {Zoccali}}]{Pasquini2003}
{Pasquini}, L., {Alonso}, J., {Avila}, G., {et~al.} 2003, in Society of
  Photo-Optical Instrumentation Engineers (SPIE) Conference Series, Vol. 4841,
  Instrument Design and Performance for Optical/Infrared Ground-based
  Telescopes, ed. M.~{Iye} \& A.~F.~M. {Moorwood}, 1682--1693

\bibitem[{{Pasquini} {et~al.}(2011){Pasquini}, {Mauas}, {K{\"a}ufl}, \&
  {Cacciari}}]{Pasquini2011}
{Pasquini}, L., {Mauas}, P., {K{\"a}ufl}, H.~U., \& {Cacciari}, C. 2011, \aap,
  531, A35

\bibitem[{{Piotto} {et~al.}(2012){Piotto}, {Milone}, {Anderson}, {Bedin},
  {Bellini}, {Cassisi}, {Marino}, {Aparicio}, \& {Nascimbeni}}]{Piotto2012}
{Piotto}, G., {Milone}, A.~P., {Anderson}, J., {et~al.} 2012, \apj, 760, 39

\bibitem[{{Piotto} {et~al.}(2015){Piotto}, {Milone}, {Bedin}, {Anderson},
  {King}, {Marino}, {Nardiello}, {Aparicio}, {Barbuy}, {Bellini}, {Brown},
  {Cassisi}, {Cool}, {Cunial}, {Dalessandro}, {D'Antona}, {Ferraro}, {Hidalgo},
  {Lanzoni}, {Monelli}, {Ortolani}, {Renzini}, {Salaris}, {Sarajedini}, {van
  der Marel}, {Vesperini}, \& {Zoccali}}]{Piotto2015}
{Piotto}, G., {Milone}, A.~P., {Bedin}, L.~R., {et~al.} 2015, \aj, 149, 91

\bibitem[{{Piotto} {et~al.}(2013){Piotto}, {Milone}, {Marino}, {Bedin},
  {Anderson}, {Jerjen}, {Bellini}, \& {Cassisi}}]{Piotto2013}
{Piotto}, G., {Milone}, A.~P., {Marino}, A.~F., {et~al.} 2013, \apj, 775, 15

\bibitem[{{Prantzos} \& {Charbonnel}(2006)}]{PrantzosCharbonnel2006}
{Prantzos}, N. \& {Charbonnel}, C. 2006, \aap, 458, 135

\bibitem[{{Prantzos} {et~al.}(2007){Prantzos}, {Charbonnel}, \&
  {Iliadis}}]{Prantzos2007}
{Prantzos}, N., {Charbonnel}, C., \& {Iliadis}, C. 2007, \aap, 470, 179

\bibitem[{{Ram{\'{\i}}rez} \& {Mel{\'e}ndez}(2005)}]{RamirezMelendez2005}
{Ram{\'{\i}}rez}, I. \& {Mel{\'e}ndez}, J. 2005, \apj, 626, 465

\bibitem[{{Renzini} {et~al.}(2015){Renzini}, {D'Antona}, {Cassisi}, {King},
  {Milone}, {Ventura}, {Anderson}, {Bedin}, {Bellini}, {Brown}, {Piotto}, {van
  der Marel}, {Barbuy}, {Dalessandro}, {Hidalgo}, {Marino}, {Ortolani},
  {Salaris}, \& {Sarajedini}}]{Renzini2015}
{Renzini}, A., {D'Antona}, F., {Cassisi}, S., {et~al.} 2015, \mnras, 454, 4197

\bibitem[{{Richter} {et~al.}(1999){Richter}, {Hilker}, \&
  {Richtler}}]{Richter1999}
{Richter}, P., {Hilker}, M., \& {Richtler}, T. 1999, \aap, 350, 476

\bibitem[{{Roediger} {et~al.}(2014){Roediger}, {Courteau}, {Graves}, \&
  {Schiavon}}]{Roediger2014}
{Roediger}, J.~C., {Courteau}, S., {Graves}, G., \& {Schiavon}, R.~P. 2014,
  \apjs, 210, 10
  
\bibitem[{{Schaerer} \& {Charbonnel}(2011)}]{SchaererCharbonnel2011}
{Schaerer}, D. \& {Charbonnel}, C. 2011, \mnras, 413, 2297

\bibitem[{{Sills} \& {Glebbeek}(2010)}]{Sills2010}
{Sills}, A. \& {Glebbeek}, E. 2010, \mnras, 407, 277

\bibitem[{{Skrutskie} {et~al.}(2006){Skrutskie}, {Cutri}, {Stiening},
  {Weinberg}, {Schneider}, {Carpenter}, {Beichman}, {Capps}, {Chester},
  {Elias}, {Huchra}, {Liebert}, {Lonsdale}, {Monet}, {Price}, {Seitzer},
  {Jarrett}, {Kirkpatrick}, {Gizis}, {Howard}, {Evans}, {Fowler}, {Fullmer},
  {Hurt}, {Light}, {Kopan}, {Marsh}, {McCallon}, {Tam}, {Van Dyk}, \&
  {Wheelock}}]{Skrutskie2006}
{Skrutskie}, M.~F., {Cutri}, R.~M., {Stiening}, R., {et~al.} 2006, \aj, 131,
  1163

\bibitem[{{Sneden}(1973)}]{Sneden1973}
{Sneden}, C.~A. 1973, PhD thesis, THE UNIVERSITY OF TEXAS AT AUSTIN.

\bibitem[{{Stetson}(2000)}]{Stetson2000}
{Stetson}, P.~B. 2000, \pasp, 112, 925

\bibitem[{{Stetson}(2005)}]{Stetson2005}
{Stetson}, P.~B. 2005, \pasp, 117, 563

\bibitem[{{VandenBerg} {et~al.}(2013){VandenBerg}, {Brogaard}, {Leaman}, \&
  {Casagrande}}]{VandenBerg2013}
{VandenBerg}, D.~A., {Brogaard}, K., {Leaman}, R., \& {Casagrande}, L. 2013,
  \apj, 775, 134

\bibitem[{{Ventura} \& {D'Antona}(2011)}]{Ventura2011}
{Ventura}, P. \& {D'Antona}, F. 2011, \mnras, 410, 2760

\bibitem[{{Ventura} {et~al.}(2001){Ventura}, {D'Antona}, {Mazzitelli}, \&
  {Gratton}}]{Ventura2001}
{Ventura}, P., {D'Antona}, F., {Mazzitelli}, I., \& {Gratton}, R. 2001, \apjl,
  550, L65

\bibitem[{{Ventura} {et~al.}(2013){Ventura}, {Di Criscienzo}, {Carini}, \&
  {D'Antona}}]{Ventura2013}
{Ventura}, P., {Di Criscienzo}, M., {Carini}, R., \& {D'Antona}, F. 2013,
  \mnras, 431, 3642

\bibitem[{{Villanova} {et~al.}(2012){Villanova}, {Geisler}, {Piotto}, \&
  {Gratton}}]{Villanova2012}
{Villanova}, S., {Geisler}, D., {Piotto}, G., \& {Gratton}, R.~G. 2012, \apj,
  748, 62
  
\bibitem[{{Wang} {et~al.}(2016){Wang}, {Primas}, {Charbonnel}, {Van der
  Swaelmen}, {Bono}, {Chantereau}, \& {Zhao}}]{Wang2016}
{Wang}, Y., {Primas}, F., {Charbonnel}, C., {et~al.} 2016, \aap, 592, A66

\end{thebibliography}

\end{document}